%% file: main.tex
\documentclass[conference]{IEEEtran}
\input{packages.tex}
\input{cmds.tex}

\input{shorthand.tex}
\IEEEoverridecommandlockouts
\def\BibTeX{{\rm B\kern-.05em{\sc i\kern-.025em b}\kern-.08em
    T\kern-.1667em\lower.7ex\hbox{E}\kern-.125emX}}
\begin{document}

\title{\system: Autonomous Hardware Resource Assignment for DNN Accelerators using Reinforcement Learning
}

\author{\IEEEauthorblockN{Sheng-Chun Kao}
\IEEEauthorblockA{Electrical and Computer Engineering\\
Georgia Institute of Technology \\
Atlanta, GA \\
felix@gatech.edu}
\and
\IEEEauthorblockN{Geonhwa Jeong}
\IEEEauthorblockA{Computer Science\\
Georgia Institute of Technology \\
Atlanta, GA \\
geonhwa.jeong@gatech.edu}
\and
\IEEEauthorblockN{Tushar Krishna}
\IEEEauthorblockA{Electrical and Computer Engineering\\
Georgia Institute of Technology \\
Atlanta, GA \\
tushar@ece.gatech.edu}
}

\maketitle

\input {\SECDIR/00-Abstract.tex}

\input {\SECDIR/01-Introduction.tex}
\input {\SECDIR/02-Background.tex}
\input {\SECDIR/03-Methodology.tex}
\input {\SECDIR/04-Evaluation.tex}
\input {\SECDIR/05-Relatedwork.tex}
\input {\SECDIR/06-Conclusion.tex}
\input {\SECDIR/07-Acknowledgment.tex}

\bibliographystyle{IEEEtranS}
\bibliography{main}

\end{document}

%% file: packages.tex
\usepackage{microtype}
\usepackage{graphicx}
\usepackage{subfigure}
\usepackage{booktabs} 
\usepackage{mathtools}
\usepackage{comment}
\usepackage[table,xcdraw]{xcolor}
\usepackage{multirow}
\usepackage{cite}
\usepackage{amsmath,amssymb,amsfonts}
\usepackage{algorithmic}
\usepackage{textcomp}
\usepackage{xcolor}
\usepackage{xspace}
\usepackage{fancyhdr}
\usepackage[hyphens]{url}
\usepackage{hyperref}
\usepackage{caption}

%% file: cmds.tex
\newcommand{\PCignore}[1]{}

\newcommand{\squishlist}{
 \begin{list}{$\bullet$}
  { \setlength{\itemsep}{0pt}
     \setlength{\parsep}{3pt}
     \setlength{\topsep}{3pt}
     \setlength{\partopsep}{0pt}
     \setlength{\leftmargin}{1.5em}
     \setlength{\labelwidth}{1em}
     \setlength{\labelsep}{0.5em} } }

\newcommand{\squishlisttwo}{
 \begin{list}{$\bullet$}
  { \setlength{\itemsep}{0pt}
     \setlength{\parsep}{0pt}
    \setlength{\topsep}{0pt}
    \setlength{\partopsep}{0pt}
    \setlength{\leftmargin}{2em}
    \setlength{\labelwidth}{1.5em}
    \setlength{\labelsep}{0.5em} } }

\newcommand{\squishend}{
  \end{list}  }

%% file: shorthand.tex
\newcommand{\system}{ConfuciuX\xspace}

\newcommand{\bayes}{Bayesian\xspace}

\newcommand{\env}{Env\xspace}
\newcommand{\eval}{eval\xspace}

\newcommand{\mobilenet}{MobileNet-V2\xspace}
\newcommand{\resnet}{ResNet-50\xspace}

\newcommand{\mnasnet}{MnasNet\xspace}

%% file: sections/00-Abstract.tex
\begin{abstract}
DNN accelerators provide efficiency by leveraging reuse of activations/weights/outputs during the DNN computations to reduce data movement from DRAM to the chip. The reuse is captured by the accelerator's dataflow. 
While there has been significant prior work in exploring and comparing various dataflows,
the strategy for assigning on-chip hardware resources (i.e., compute and memory)
given a dataflow that can optimize for performance/energy while meeting platform constraints of area/power for DNN(s) of interest is still relatively unexplored.
The design-space of choices for 
balancing compute and memory 
explodes combinatorially, as we show in this work (e.g., as large as $O(10^{72})$ choices for running \mobilenet), making it infeasible to do manual-tuning via exhaustive searches. It is also difficult to come up with a specific heuristic given 
that different DNNs and layer types exhibit 
different amounts of reuse.

In this paper, we propose an autonomous strategy called \system to find optimized HW resource assignments for a given model and dataflow style. \system leverages a reinforcement learning method, REINFORCE, to guide the search process, leveraging a detailed HW performance cost model within the training loop to estimate rewards.
We also augment the RL approach with a genetic algorithm for further fine-tuning.
\system demonstrates 
the highest sample-efficiency for training compared to 
other techniques such as \bayes optimization, 
genetic algorithm, simulated annealing, and other RL methods.
It
converges to the optimized
hardware configuration 4.7 to 24 times faster than alternate techniques. 

\end{abstract}




%% file: sections/01-Introduction.tex
\section{Introduction}
\label{sec:intro}


Deep neural networks (DNNs) are being deployed into many real-time applications such as autonomous driving, mobile VR/AR, and recommendation systems. However, DNNs are often strictly constrained by end-to-end latency or energy.
This has opened up extensive 
research on computationally
efficient DNN models~\cite{sandler2018mobilenetv2,tan2019efficientnet} 
and hardware accelerators~\cite{tpu, chen2016eyeriss, du2015shidiannao, nvdla, maeri}. 


\input{figure/raman_fig}
The architecture of DNN accelerators is determined 
by two key components: \textit{dataflow} style and total \textit{HW resources}.
The dataflow comprises 
the computation order, parallelization-strategy, and tiling strategy employed by the accelerator~\cite{maestro, chen2016eyeriss}.
The HW resources comprise of the total on-chip compute (hereby referred to as ``PEs") and on-chip memory (hereby referred to as ``Buffers"). The underlying network on chip (NoC) bandwidth, and the corresponding implementation~\cite{chen2016eyeriss, maeri} and area depends on dataflow-style and assigned HW resources.
For the same dataflow strategy, multiple 
resource assignments are possible, as shown in \autoref{fig:raman_fig}.
The dataflow and/or the HW resources are either fixed at design-time (which is the common-case), or can be tuned at compile time 
(if the accelerator is reconfigurable, such as CGRA-based~\cite{maeri} or FPGA~\cite{cong_fpga}).

A lot of previous research has focused on designing efficient dataflow 
strategies to extract reuse.
For e.g., NVDLA \cite{nvdla}, Eyeriss \cite{chen2016eyeriss}, and ShiDianNao \cite{du2015shidiannao} are examples of DNN 
accelerators that employ different dataflow strategies. 
Frameworks like MAESTRO~\cite{maestro} and Timeloop~\cite{timeloop} exist to 
contrast the performance benefits
of various dataflows.
Most accelerators choose 
a dataflow strategy
based on the expected dimensions and shapes of the DNNs they will run.
For example, the NVDLA~\cite{nvdla} dataflow keeps weights stationary at PEs, and parallelizes 
across 
input channels and output channels, as shown in \autoref{fig:raman_fig}, optimizing for mid and late layers of many CNNs like ResNet~\cite{resnet} that exhibit this property.
The Eyeriss~\cite{chen2016eyeriss} dataflow 
parallelizes across the 
activation and filter rows,
and keeps filter rows 
stationary at the PEs.
Reconfigurable accelerators like MAERI~\cite{maeri} allow the dataflow strategy to be configured for every layer~\cite{mrna}.

\input{figure/layer_whole_def}

Given a dataflow, the assignment of HW resources is the next crucial
part of the DNN accelerator design process.
In fact, for the same dataflow, different choices for HW resources can lead to drastically different latency and energy for a given DNN, as we show later in \autoref{fig:motivation}. Some recent studies have shown that HW resource assignment plays a more important role in determining 
the accelerators' performance than its dataflow \cite{yang2018dnn}.
However, determining the 
policy for assigning HW resources is still very much an open problem, with
prior works on HW Design-Space Exploration almost exclusively relying on exhaustive searches~\cite{shen2017maximizing,li2016high,zhang2016energy,jiang2018heterogeneous,jiang2019xfer,zhang2018dnnbuilder,chen2019cloud,wei2018tgpa,maestro,cong_fpga}.

The focus of this work is on the aforementioned HW resource assignment problem.
The HW resource assignment depends on how they will be used by the DNN during runtime.
We consider two deployment scenarios in this work, shown in \autoref{fig:layer_whole_def}.
Layer Sequential (LS) involves mapping and running the DNN layer by layer
on the accelerator, while Layer Pipelined (LP) maps and runs the entire DNN model over the accelerator.
The LS approach is typically leveraged in cloud settings for larger models~\cite{resnet} that 
do not fit on-chip, while the LP approach is popular when running smaller optimized models~\cite{sandler2018mobilenetv2, tan2019mnasnet, tan2019efficientnet} on
IoT devices. 

The HW resource assignment problem 
is an optimization problem where 
the design goal is to achieve an objective such as minimum end-to-end latency or energy,
while meeting some platform (IoT/Cloud) constraints such as maximum power or chip area.
The design-space of 
valid solutions is non-trivial.
Consider an LP deployment; suppose we have a total of $P$ PEs and $B$ buffers that fit within the area/power budget and need to be divided among $N$ layers of the DNN.
Assuming each layer gets at least one PE and one buffer, the number of combinations for PEs and buffers is ${P-1 \choose N}$ and ${B-1 \choose N}$ respectively~\cite{stars_and_bars}.
This makes the total possible design choices ${P-1 \choose N}$$\times$${B-1 \choose N}$, which is $O(10^{72})$
for an accelerator with 128 PEs, 128 buffers running the 52-layer \mobilenet.
This design-space is nearly impossible to enumerate to search exhaustively for an optimum solution, as we discuss in \autoref{sec:background}.


\input{figure/RL_algo}

In this work, we develop an autonomous mechanism to efficiently search through the HW design-space.
\autoref{fig:workflow} shows an overview of our proposed workflow called \system{}.
It takes the target model, platform constraint, deployment scenario (LS or LP), and optimization objective (latency/energy) as input, 
and determines an optimized HW assignment strategy (number of PEs and buffers).
\system{} leverages reinforcement learning (RL) to perform a global coarse-grained search, followed by a genetic algorithm (GA) for fine-grained tuning.


Recently RL has been demonstrated within compilers/mappers~\cite{ahn2019reinforcement, mirhoseini2017device, spotlight,paliwal2020reinforced} for tiling and mapping DNNs over accelerators. 
\system focuses on leveraging RL for exploring the search space during accelerator design.


We evaluate \system on popular DNN models, including \mobilenet \cite{sandler2018mobilenetv2}, \resnet \cite{resnet}, and \mnasnet \cite{mnasnet},
GNMT~\cite{gnmt}, Transformer~\cite{vaswani2017attention} and NCF~\cite{he2017neural}. We evaluate our HW resource assignment method under three  dataflow styles, NVDLA-style, Eyeriss-style, and ShiDianNao-style\footnote{We refer to them as -style since we follow the dataflow behavior part of these accelerators but allow flexibility in varying free dimensions (number of PEs and tile-sizes) at design/compile time.}. We evaluate both cloud and IoT device platform constraint setting. 
We also demonstrate a joint search for dataflow and HW assignments.
We contrast our approach against other  optimization mechanisms including Genetic Algorithm \cite{ga}, simulated annealing \cite{SA}, \bayes optimization \cite{bayes} and other state-of-the-art RL algorithms \cite{ddpg, td3, sac, acktr,a2c,ppo2} and observe that it consistently outperforms alternate schemes both in terms of solution quality (latency/energy) and search time (4.7 to 24 $\times$ faster). 

The paper is organized as follows. \autoref{sec:background} provides relevant background on DNN accelerators and relevant optimization methods; \autoref{sec:methodology} describes \system{} in detail; \autoref{sec:experiments} presents comprehensive evaluations; \autoref{sec:relatedworks} 
presents related work and \autoref{sec:conclusion} concludes.

%% file: figure/raman_fig.tex
\begin{figure}[!t]
\begin{center}
\includegraphics[width=1\linewidth]{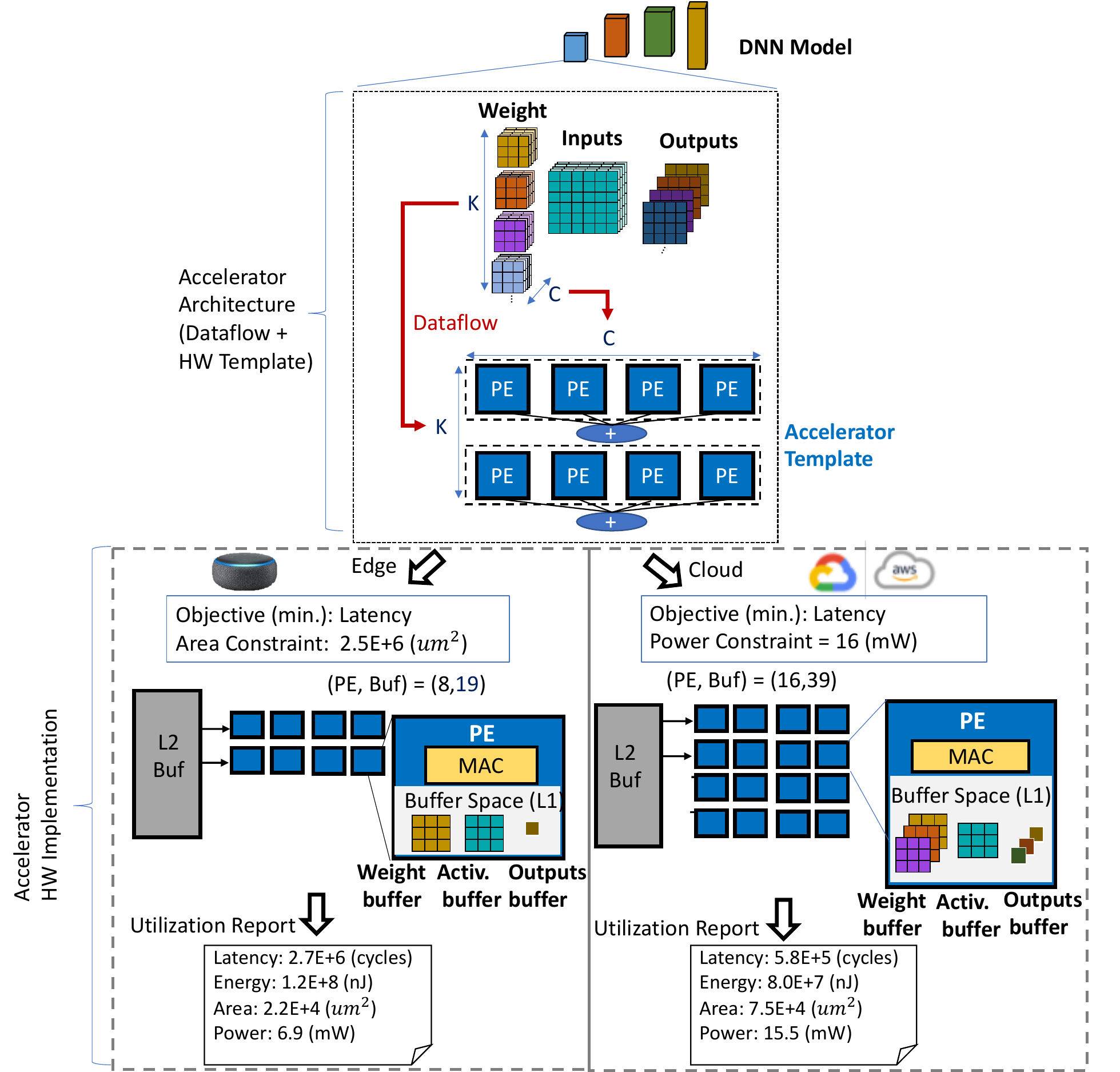}
\end{center}
\caption{Different HW resource combinations with the same NVDLA-style dataflow.}

\label{fig:raman_fig}
\end{figure}

%% file: figure/layer_whole_def.tex
\begin{figure}
\begin{center}
\includegraphics[width=1\linewidth]{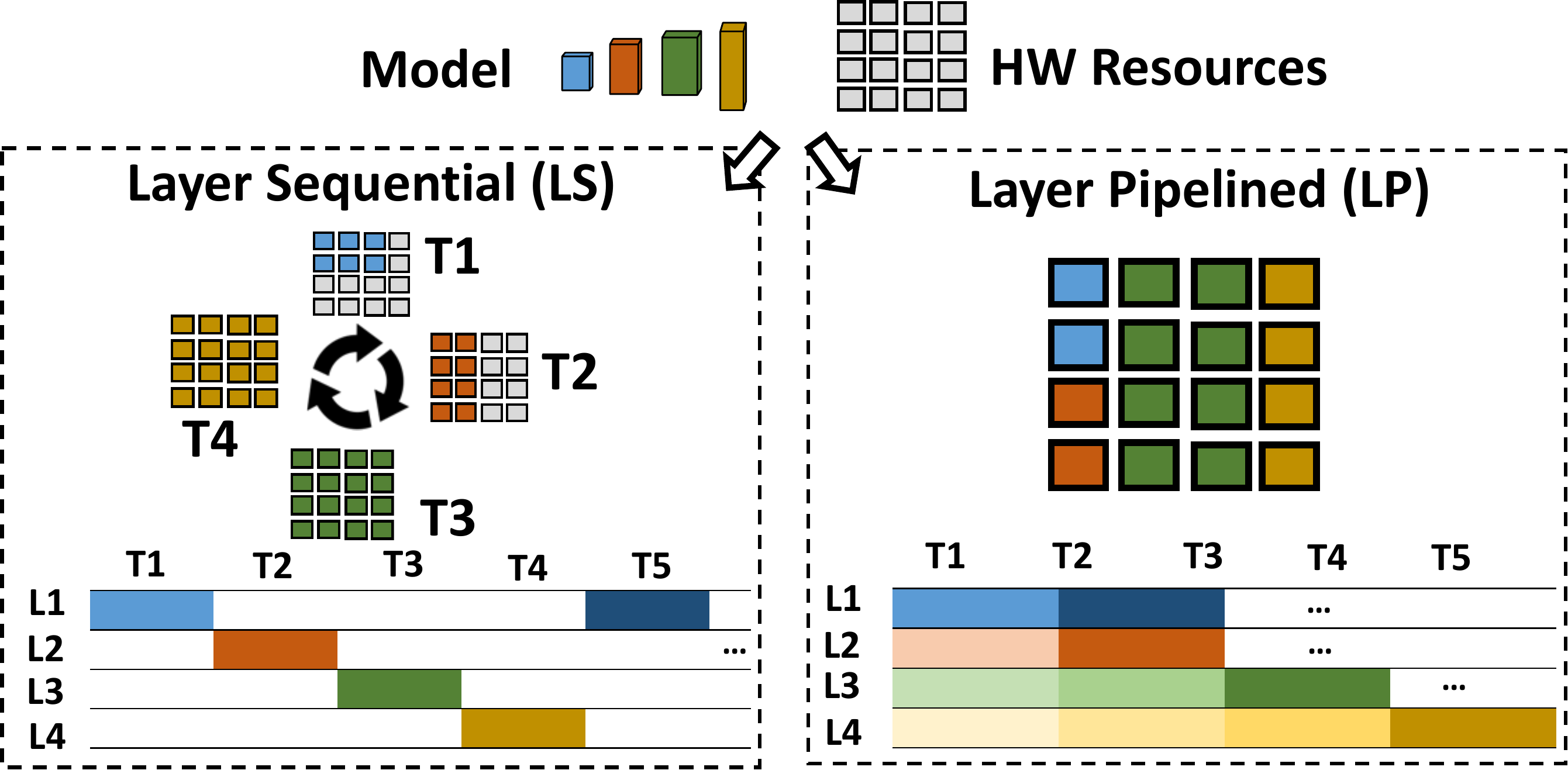}
\end{center}

\caption{ DNN deployment scenarios and corresponding HW assignments. In Layer Sequential (LS), 
each layer of the model is mapped one by one on the entire accelerator, with all on-chip compute and memory assigned to it;
in Layer Pipelined (LP), the entire model is mapped and run in a pipelined manner, with the compute and memory partitioned across all layers. }

\label{fig:layer_whole_def}
\end{figure}

%% file: figure/RL_algo.tex
\begin{figure*}
\begin{center}
\includegraphics[width=1\linewidth]{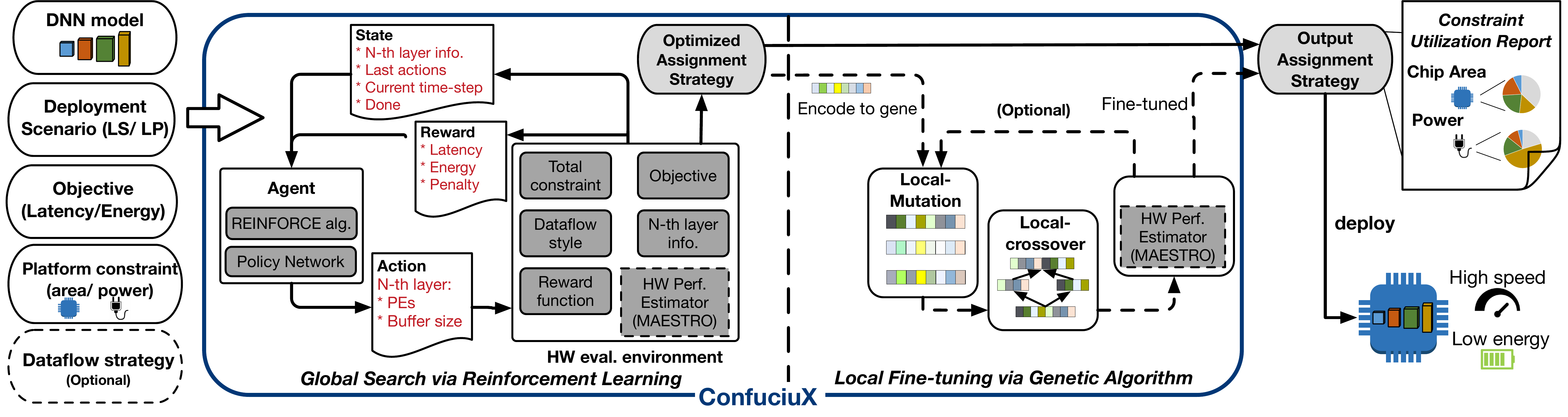}
\end{center}

\caption{Overview of \system.}
\vspace{-4mm}

\vspace{-0.1cm}
\label{fig:workflow}
\end{figure*}

%% file: sections/02-Background.tex
\section{Background and Motivation}
\label{sec:background}
\subsection{DNN Accelerator Architecture}
We discuss and define some key terms (highlighted in bold) that we will use throughout the paper.

\subsubsection{\textbf{Hardware Resources}}
Spatial DNN accelerators 
comprise an array 
of 
Processing Elements (called \textbf{\textit{PE}} 
in this paper), 
as shown in \autoref{fig:raman_fig}.
Each PE has 
a MAC to compute partial sums,
and local (aka ``L1")
buffers (called \textbf{\textit{Buffer}} in this paper) to store
weights, activations,
and partial sums.
The accelerators 
also house a 
global shared (aka ``L2") buffer to prefetch activations and weights from DRAM for the 
next tile of computation that will be mapped over the PEs and L1 buffers.
Networks-on-Chip (NoCs) are used to 
distribute operands from the global L2 buffer
to the L1 buffers in the PEs, and collect 
the partial or full outputs and write them back to the global L2 buffer.

\subsubsection{\textbf{Dataflow}}
The mechanism for orchestrating 
data from the global SRAM to the 
local PE buffers (i.e., computation order, parallelization strategy across PEs,
and tiling strategy) is called \textit{dataflow}.
For example, accelerators like 
NVDLA \cite{nvdla}, Eyeriss  \cite{eyeriss_isscc}, ShiDianNao \cite{du2015shidiannao}, TPU \cite{tpu} all employ unique dataflow strategies~\cite{chen2016eyeriss,maestro}.

\subsubsection{\textbf{Design-Point}}
In this work, 
we assume that
the number of PEs 
and buffers are free-variables\footnote{The number of buffers depends on the maximum tile size of 
weights/inputs/outputs that the accelerator supports in each PE~\cite{maestro}.
In this work, we control the buffer size by changing the
tile size for filters.}  that can be tuned independently in an accelerator during design-time/compile-time (depending on the deployment scenario discussed later in \autoref{sec:deployment}).
We call each combination of (PE, buffer) as a unique \textbf{\textit{design-point}} in this paper.
Given a specific dataflow, each 
design-point in-turn determines 
the size and number 
of other components within the accelerator.
For e.g., the number of PEs and L1 buffer sizes determine the minimum size of the 
global L2 buffer to hold the next tile of \textit{unique data} that will need to be sent to the PE buffers~\cite{maestro, timeloop}. The L2 can then be sized to be double this value to prefetch the next tile from DRAM while the current one is being processed. Similarly, the design-point also determines the NoC bandwidth for stall-free distribution of operands to the PEs and collection of outputs \cite{chen2016eyeriss, maeri} for that dataflow.
Thus, we choose only the PE and buffer size 
as the independent design-parameters 
in this work.
It is certainly possible to let the 
L2 size and NoC bandwidth also be 
independent parameters, 
but this could lead over-provisioning (i.e., under-utilization)
or under-provisioning (i.e., stalls)~\cite{maestro}.

\subsection{DNN Accelerator Performance and Cost Modeling}
The overall runtime, throughput
and energy-efficiency of a DNN 
accelerator depends on 
three aspects: DNN model, 
mapping (dataflow and tile sizes),
and HW resources~\cite{maestro}.
We briefly discuss this cross dependence.

\squishlist
\item \textbf{DNN Models.}
There are myriads of DNN models and most of them are built using different combinations of some common layers. Convolutional layers (2D/depth-wise/point-wise) dominate in DNNs like \resnet, \mobilenet \cite{sandler2018mobilenetv2}, and InceptionNet \cite{szegedy2017inception} targeting image processing tasks.
Fully connected layers or MLPs are often used as the last layer in many DNNs models, as hidden layers of RNNs, and in language models~\cite{gpt2} and machine translation~\cite{gnmt}.
Different layer types expose different amounts of data reuse opportunities, which 
can be exploited 
by DNN accelerators
depending on the mapping.

\item \textbf{Mapping.} A mapping~\cite{maestro,timeloop} 
refers to the dataflow and specific tile sizes.
The dataflow is the mechanism 
for reusing data across 
time (via buffers) and space (over wires).
The tile sizes 
are bound by 
the L1 and L2 
buffer sizes within the accelerator.
The total number of tiles 
depends on the DNN model size 
and the dataflow strategy.

\item \textbf{Hardware Resources.} 
The total number of PEs in the accelerator determines peak throughput, while 
the buffer sizes in each PE 
determine the amount of reuse that each PE can exploit within the 
tile of computation mapped on to it in each time iteration.
\squishend

\textbf{Cost Models.}
The interdependence between DNN model layer shape, mapping strategy 
and hardware resources is captured 
by cost models like MAESTRO~\cite{maestro} 
and Timeloop~\cite{timeloop} that 
can 
\textit{analytically} determine the reuse across time/space
and accordingly estimate
the runtime and utilization.
These cost models 
can also estimate the area and power of the accelerator for a given design-point.


\input{figure/motivation}

\subsection{DNN Model Deployment on Accelerators}
\label{sec:deployment}
We focus on two DNN model deployment scenarios illustrated in \autoref{fig:layer_whole_def}.

\textbf{Layer Sequential (LS).}
We use the same underlying architecture to run the DNN model layer-by-layer.
The specific (PE, buffer) design-point is chosen at design-time by some heuristic such 
as one that performs the best 
for most layers of the target DNNs.
Naturally, over-provisioning and under-utilization may happen for some of the layers during deployment since they favor different HW resource configuration
~\cite{chakradhar2010dynamically, farabet2010hardware,farabet2009cnp,sankaradas2009massively}. We quantify this further in \autoref{sec:experiments}. 

\textbf{Layer Pipelined (LP).}
With the advancement of technology, more computation logic can sit in a single chip. Many efficient models are being designed to fit completely onto the chip for embedded platforms \cite{tan2019efficientnet, sandler2018mobilenetv2}. 
LP maps and runs the entire DNN model over the accelerator. 
For this model, we assume an underlying accelerator can heterogeneously partition the (PE, buffer) resources at either design-time (e.g., ASIC~\cite{bang201714,yin2018141,zheng2019ultra})
or compile-time (e.g., CGRA~\cite{maeri}/FPGA~\cite{cong_fpga}).
The challenge becomes finding the optimum (PE, buffer) distribution for each layer, which is 
crucial for maximizing performance \cite{shen2017maximizing,li2016high,zhang2016energy,jiang2018heterogeneous,jiang2019xfer,zhang2018dnnbuilder,chen2019cloud,wei2018tgpa}.

\subsection{Challenge: Design-Space for HW Resource Assignment}
\label{sec:design_space}

There can be myriad design points that fit the platform power/area constraint, each with drastically different performance/energy. 
As an example, 
we visualize the fine-grained hardware design space for a DNN accelerator in \autoref{fig:motivation}, plotting the latency and energy when running three different layers of \mobilenet on different accelerator design points with NVDLA-style dataflow. 
The numbers were obtained from MAESTRO~\cite{maestro_web}.
Each point in the graphs 
is a design-point (i.e., \{number of PEs, L1 Buffer per PE\}). We sweep the PEs from 1 to 64, 
and number of filters that can  be mapped from 1 to 800 (which in turn sweeps the L1 buffer size from 6B to 2800B).
For the same design-point, the top half of \autoref{fig:motivation} shows the the number of PEs, and the bottom half shows the size of buffers.
Each design-point leads to a unique latency, energy, and area consequence.

From \autoref{fig:motivation}, we can conclude that for the same area, the range of possible latency and energy values is quite significant. This gets exacerbated when trying to find an optimal design point that works well across most layers in an LS deployment or for finding the combination of (PE, Buffer) per layer 
in an LP deployment.
As mentioned in \autoref{sec:intro}, with just 128 PEs and 128 buffers,
the design-space for \mobilenet deployment is $O(10^{72})$, making it infeasible for any HW Design-Space Exploration (DSE) to sweep exhaustively.
%
%
Most prior accelerator prototypes have picked specific design-points for their dataflow (e.g., 168 PEs in Eyeriss~\cite{eyeriss_isscc} and 64 PEs in ShiDianNao~\cite{du2015shidiannao}) without exploring the design-space across different area/power constraints. 
This is the focus of this work.

\subsection{Optimization Methods for Design-Space Exploration}
The following optimization methods exist today for architects to perform Design-Space Exploration (DSE) and form our baselines.

\textbf{Exhaustive search} will lead to a global optimum, but is nearly impossible to sweep for vast design spaces.
\textbf{Grid search} is an exhaustive search with a coarse-grain sampling step, which makes the process approachable. 

\textbf{Random search} randomly samples design points in an unknown search space and keeps the best solution.
It has been shown to be competitive for optimization problems in various fields~\cite{bergstra2012random,larochelle2007empirical,pinto2009high,salimans2017evolution}.

\textbf{Simulated annealing \cite{SA}} adds an exploitation step to random search (which is always exploring).
It randomly samples and accepts points that improve the objective, but also with a certain probability accepts points that may worsen the objective. The probability is controlled by a hyper-parameter \textit{temperature}. Higher temperature will increase the probability to accept worse points, causing more randomness, and vice versa. Simulated annealing is used for compiler optimization for CPU and software \cite{zhong2009tuning} and also tile and loop scheduling for DNN workload \cite{chen2018tvm}.
 
\textbf{Genetic Algorithm (GA) \cite{ga}} is a method where we encode the dimension of each design point as a gene. With all the dimensions specified, a design-point is called a genome. We initialize the algorithm with several randomly sampled design points (genomes) and these genomes form a generation. Then we evaluate the fitness of individuals of this generation. We keep well-performing individuals and use them to reproduce the next generation with mutation and crossover. Generation by generation, GA will converge to an optimized point. STOKE \cite{schkufza2013stochastic} and TensorComprehensions \cite{vasilache2018tensor} use GA to search the space of DNN code optimization.

\textbf{\bayes optimization}
\cite{bayes} builds a surrogate for the objective and quantifies the uncertainty in that surrogate using a Bayesian machine learning technique. The optimization method can be constructed by the Gaussian process, Random forest, or Tree Parzen Estimator. It selects the next values to evaluate by applying criteria to the surrogate. It evolves the surrogate model and sampling criterion simultaneously. The concept is to limit the evaluation of the objective function by spending more time choosing the next values for sample efficiency.
Some works use \bayes optimization to search for DNN hyper-parameters \cite{stamoulis2018hyperpower, stamoulis2018designing,parsa2019pabo}.

\subsection{Reinforcement Learning for Design-Space Exploration}
Reinforcement Learning (RL) algorithms are often used 
in games~\cite{torrado2018deep, mnih2013playing} as they 
are useful in sequential decisions.
More formally, this is a 
Markov decision process (MDP).
In this work, we show that 
determining the
appropriate number of 
PEs and buffers for 
a series of DNN layers 
to minimizes the overall platform latency/energy while staying within an area or power budget
can be viewed as a MDP.
Therefore we find RL algorithms to be a promising approach for this problem to increase sample efficiency of the search, compared to 
baseline optimization methods
that use no information 
from the current state.

\textbf{Reinforcement Learning Terminology.}
The goal of an RL agent is to 
continuously interact with an \textbf{environment},
observe the current \textbf{state},
take one or more \textbf{actions},
observe the \textbf{reward} from the environment,
and update its underlying 
\textbf{policy network}.
With time, the policy network 
learns to predict actions that 
can maximize reward.
We discuss our RL-based HW resource exploration next.

%% file: figure/motivation.tex
\begin{figure*}
\begin{center}
\includegraphics[width=1\linewidth]{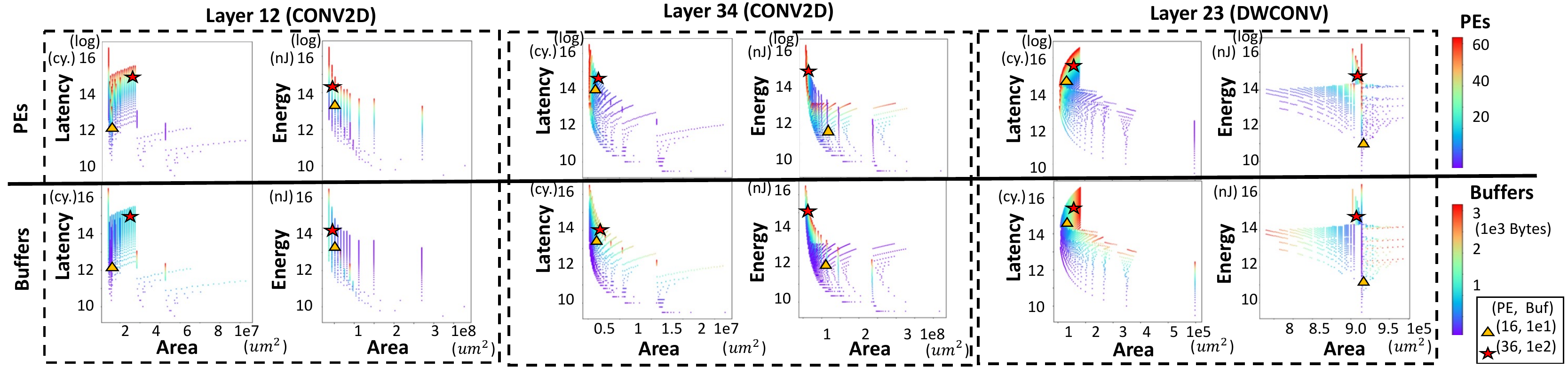}
\end{center}

\caption{Hardware design space for an accelerator running a few example layers from \mobilenet using NVDLA-style dataflow in the MAESTRO cost model~\cite{maestro}. Each dot represents a design-point (number of PEs,  L1 buffer per PE). For the same design-point, top half plots the number of PEs and bottom half plots the size of L1 buffers (in bytes). The corresponding latency, energy, and area are shown. 
Star and triangle icons represent two same design-points, and highlight how how their performance changes across different layers. 
The design-space for a given area-budget and/or latency/energy target is extremely large, and no design-point is optimal for all cases.  (DWCONV: Depth-wise CONV.)}
\vspace{-0.4cm}
\label{fig:motivation}
\end{figure*}



%% file: sections/03-Methodology.tex
\section{\system}
\label{sec:methodology}

In this work, we cast the DNN accelerator resource assignment DSE challenge as a reinforcement learning (RL) problem using REINFORCE~\cite{reinforce} for a global search, followed by a GA for local fine-tuning.
\autoref{fig:workflow} 
demonstrates the workflow of \system.
We provide a high-level overview next.
The inputs to \system are the target DNN model, the deployment scenario, the optimizing objective, and the platform constraints, and it outputs an optimized HW resource assignment strategy, as shown in ~\autoref{fig:workflow}.
The first stage of \system trains a RL agent to recommend 
an optimized assignment of PEs and buffers for a target DNN, platform constraint, and deployment scenario.
The agent is trained by having it continuously
generate resource assignments as ``actions" which are evaluated by a
detailed but fast analytical model for DNN accelerators called 
MAESTRO \cite{maestro_web} that acts as the environment (Env).
The ``rewards" output by the environment are used to train the underlying policy network, with the aim of maximizing the reward.
We incorporate platform constraints (area/power) as inputs to the environment to punish actions that violate the constraints.
To speed up the search process, the RL agent searches through the HW assignments in a coarse-grained manner. 
Once it converges to an optimized strategy, we fine-tune the assignment further using GA.
We discuss the various 
components of \system next.

\subsection{RL Agent}
In our system, the RL agent processes the 
target DNN model in a layer-wise manner.
We term the whole process (episode in RL parlance) as an epoch.
We treat each layer as a different time-step.
At each time-step, the agent makes two actions per-layer: the number of PEs and Buffers. 
It interacts with the environment to collect 
the rewards.
We feed the rewards, along with the previous layer's actions to the policy function, to help the agent optimize sequential decisions. 
The policy network gets updated at the end of each epoch. An epoch terminates when the agent fails the constraint or successfully made $2N$ actions for a $N$-layer model.

\subsubsection{\textbf{Choice of RL Algorithm: REINFORCE}}
Modern RL algorithms~\cite{ppo2, ddpg, td3, sac, a2c, acktr} typically use two underlying neural networks - an ``actor" and a ``critic". The actor formulates the policy for taking actions while the critic approximates the value function that predicts expected reward to help the training of policy.
We experimented with a suite of RL algorithms for \system{} and found that REINFORCE \cite{reinforce} works best. We show these 
results in \autoref{subsec:compare_rl}.
REINFORCE only has an actor network (no critic), and updates its underlying policy network directly using rewards from the Env.
Since the design space of HW resource assignments is extremely discrete and irregular, we observed that RL algorithms with critic networks fail to approximate the value function accurately and in turn disturb the policy learning process. We show this later in \autoref{subsec:compare_rl}.

\subsubsection{\textbf{Policy Network Architecture}}
The policy network in REINFORCE is a neural network tasked to learn the policy to maximize the probability of receiving better reward. 
We use an RNN as the policy network with one LSTM hidden layer of size 128.
The reasoning behind an RNN-based network is as follows. We impose a hard constraint on the overall area (or power) consumption. 
Each action (PE, buffer) adds to area/power. Thus, any future action should depend on the previous action. The recurrent connections in the RNN capture this relationship and learn the constraint.
We implemented and evaluated both RNN-based and MLP-based policy networks and provide a quantitative analysis in \autoref{subsec:policy_network}.



\subsection{Observation (State)}
 We construct a 10 dimensional observation space. At $t^{th}$ time step, the observation ($O_t$) is expressed as follows
\begin{align}
\hskip\parindent & \begin{gathered}
    O_{t}=(K_{t}, C_{t}, Y_{t}, X_{t}, R_{t}, S_{t}, T_{t}, A_{t}^{PE}, A_{t}^{Buffer}, t)
    \end{gathered} 
\label{formula:O}
\end{align}
The layer shape (assuming convolutions) results in the first 7 dimensions\footnote{for other layers like MLP/GEMM, we use three dimensions (M,N,K) to describe the (M,K), (K,N) and (M,N) matrices.}. $K_{t}$ and $C_{t}$ are number of output and input channels.  $Y_{t}$ and $X_{t}$ are the size of Y and X axis of the input activations. $R_{t}$ and $S_{t}$ are the size of Y and X axis of the weight kernel. $T_{t}$ is the indicator of layer type such as CONV or DWCONV (Depth-wise CONV). $A_{t}^{PE}$ and $A_{t}^{Buffer}$ are the actions of the previous layer which we feed into the RNN controller. The last dimension $t$ indicates $t^{th}$ layer.
Finally, we normalize all the dimensions of observation to the range of [-1, 1] to stabilize the training.

\subsection{Action Space}
At each time step, the agent makes an action pair (PE, Buffer), which formulates the action space.
To efficiently step through the huge design space (\autoref{sec:design_space}), the RL agent uses coarse-grained steps to navigate through it.
In particular, we use $L=12$ different values for the PEs and Buffers, as shown in \autoref{table:action_value}.
We demonstrate the effect of $L$ later in \autoref{subsec:policy_network}.
The specific values for PE at each level are chosen by the marginal observed return of HW performance to the number of PEs. For example, increasing PE from 1 to 2 could potentially double the HW performance, while increasing PE from 64 to 65 would provide slight or mostly no improvement.
We choose the Buffer size value at each level according to the input of dataflow-style at design time. In NVDLA-style, with 3$\times$3 weight as an example, we dispatch the computation to each PE along K dimension (\autoref{fig:raman_fig}). 
Each PE would receive $k$ number of 3$\times$3 weights, 3$\times$3 corresponding inputs, and generate $k$ number of outputs, which makes the buffer size $9\times k+9\times 1+1\times k$, where k=1, 2,..., 12, as shown in \autoref{table:action_value}.
Note that once the RL agent converges, \system uses fine-grained steps using GA to get to an optimized configuration, as described later in \autoref{subsec:our_ga}.



\input{tables/action_values}

\subsection{Platform Constraint and Objective}
Each action pair (PEs, Buffers) defines the per-layer power/area constraint consumption and the per-layer energy/latency cost, which are our optimization targets. The goal of the accelerator design process is to 
optimize the cost for running the entire model,
while meeting the platform constraint.

\textbf{Constraint.}
The accelerator is constrained by the budget of the targeted platform. We consider two categories of constraints: power and chip area. We have full flexibility to design architecture such as assigning a different number of PEs and Buffers or changing dataflow-style, as long as the design meets the constraint. In this paper, we evaluate power and area constraints across cloud and IoT platforms, as described in \autoref{sec:methodology}. 

\textbf{Objective.}
We evaluate two design objectives in this work: minimum overall latency and minimum overall energy cost when the entire model is run on the accelerator, either via LS or LP. Other objectives can also be considered (say EDP or Power/Area for instance). While approaching the objective, the design should always fit the platform constraint. Minimizing the latency and energy is a non-trivial task since their dependence on the number of PEs and Buffers is not straight-forward. For e.g., increasing PEs would increase the level of parallelism; however, it would also increase the number of fetched data, which could potentially increase the latency. As for energy, increasing PEs and Buffers would increase the power; however, it could potentially decrease the energy because of the shorter execution time. 

\textbf{Co-optimization of layers.}
The RL agent is trained to be aware of the platform constraints. The agent should learn to optimize the resource assignment for each layer and the allocation of the constraint budget at hand to each layer simultaneously. We use RNN as the backbone of RL agent to enable it to memorize its entire epoch of decisions so that it could be aware of the consumption of the total budget. The Env checks the budget that is still left ($L^{budget}$) at every time step, and penalizes the RL agent once it is violated.

\subsection{Reward Function}
\textbf{Reward.}
Since we are executing in a sparse reward domain, where the performance is only given at the end of the episode, we train the agent with a temporal layer-wise performance feedback for reward shaping. The sum of the layer-wise performance does not directly indicate the final entire model performance, which is our objective. However, it guides the RL agents.

We construct the reward function $R$ as follows.
\begin{align}
\hskip\parindent&
    R=
    \begin{cases}
 &  P_{t} - P^{min}\text{, if  }  L^{budget} \geq 0 \\ 
 & Penalty\text{, otherwise}
\end{cases}
\label{formula:R}
\end{align}
$P_{t}$ is the HW performance\footnote{We use the term performance for generality. It could be latency, or energy, or any other objective we are minimizing.} of the current layer. $P^{min}$ is the current lowest layer-wise performance across all time-steps and all epochs.
This is tracked during the training process.

We find that the $P^{min}$ term stabilizes the training. The insight behind it is as follows. 
First, as shown in \autoref{fig:raman_fig}, the reward value for HW performance, such as number of cycles, can be extremely large, which can make the relative improvement seem insignificant across epochs.
Thus, keeping a $P^{min}$ across all epochs emphasizes the relative difference. 
Second, the term $P^{min}$ makes the reward always positive while the platform constraint is not violated, which makes the RL agent easier to learn from positive reward and negative penalty. 

\textbf{Penalty.}
We penalize the RL agent when the resource constraint is violated. To teach the RL agent to forbid the failing point with reasonable penalty, we accumulate all the rewards experiences in this episode, and use negative of the accumulated value as a penalty. The reason is that the range of reward for different HW performance (latency, energy) can have an order of magnitude difference. Therefore, a threshold-based constant penalty ~\cite{openai,popov2017data,wu2016training}, which is usually applied, is not feasible. Also, we need the penalty that is at the correct scale so that it is large enough to penalize the agent and small enough to not deviate the learned policy too much once bad decision is made.

At the end of the episode, we normalize rewards in each time step to standard distribution and use the standardized reward to train the agent. We also apply a discount factor ($d$). We empirically found $d=0.9$ is a generic good default value for this problem. 

\subsection{Interactive Environment (\env)}
\textbf{Structure.}
The \env is initialized with the target model(s), dataflow, platform  constraint, and the optimizing objective (latency/energy).
Env tracks the consumed constraints of each time step and the $P^{min}$ across all episodes.

\textbf{HW performance estimator (\eval).}
We use MAESTRO \cite{maestro_web}, an open-source DNN accelerator microarchitectural model, to determine the performance of each accelerator design-point during the training process.
MAESTRO takes the DNN model, dataflow-style, and HW configuration as an input.
Internally, it estimates all possible reuse opportunities for the given dataflow and HW resources,
and estimates statistics such as latency, energy, runtime, power, and area.
MAESTRO's HW model assumes a  spatial DNN accelerator with PEs, L1 buffers, a shared L2 buffer, and an NoC between the buffers. It can support any dataflow (specified via a data-centric DSL~\cite{maestro}).
The number of PEs is an input parameter, while the L1 and L2 buffer sizes are estimated based on the tile-sizes for the dataflow.
It supports both layer-wise and model-wise evaluation.

\subsection{Local fine-tuning using GA}
\label{subsec:our_ga}
We use a two-stage optimization to search for a fine-grained solution, as shown in \autoref{fig:workflow}. The first and major part is the RL based coarse-grained global search. The second is the Genetic Algorithm (GA) based fine-grain local search. We use two-stage optimization for efficiency, since increasing the level of actions, $L$ by 1 would increase the design space by $(\frac{L+1}{L})^{2N}$. Using \mobilenet and $L=12$ as an example, we would increase the design space discussed in \autoref{sec:design_space} by another 64 times.

RL shows higher sample efficiency and converges to better optimum point comparing to other optimization methods, as shown later in \autoref{sec:experiments}. GA is simple and fast, but converging to less optimum value comparing to RL or sometimes cannot converge. According to the observation of the behavior of GA, it sometimes fails to learn the constraint and optimize the objective simultaneously, leading to a great portion of populations actually violating the constraint, which pollute the genomes of the future generation. However, if we start GA with a good initialization and mutate/crossover genes carefully, which decreases the complexity of the problem, GA could reach good result. Therefore, GA becomes a good candidate as a second stage fine-tuning if we initialize it with the first-stage solution. Even though a continuous RL algorithm~\cite{ddpg, td3, sac} could be another candidate for the second stage, we find that the problem complexity of the second stage is simple that GA is adequate to tackle it.
The details of the GA algorithm are described next.

\textbf{Initialization.} 
Assuming a DNN model with $N$ layers, a design-point would include $N$ actions for PEs and $N$ actions for Buffers. We encode this design-point into a genome with $2N$ genes, 
where a gene represents an action for PE or Buffers.
We initialize the first population with the genome formulated by the solution from the first (RL) stage.

\textbf{Local mutation.} We mutate the gene locally. We only mutate the gene by a step difference of the current value. For e.g.,
for a gene representing PE=64, we could mutate it to value in the range of [60, 68] when the step is 4. 
This conservative mutation can reduce the number of invalid 
genomes, which does not conform to the constraint, and assure we have good portion of valid parents to reproduce.

\textbf{Local crossover.} The crossover of two genomes is unlikely to conform to the constraint, since it can break the learnt relationship between HW resource assignment of each layer.
For e.g., suppose we have parents A and B, both with good fitness and lie within the constraint. However, 
A tends to assign more HW resources on early layers and B tends to assign more HW resources on late layer. When we blend their genes for the next generation, the platform constraints might get violated by some children: a child with early genes from A and late genes from B may over-request HW resources for every layer, violating the constraints. Alternately, a child with early genes from B and late genes from A 
may under-request HW resources for each layer, leading to less performance.
Thus, we crossover the genome locally within a parent by exchanging genes for (PE, Buffers) between two layers of a model. In other words, we pick two pairs of genes representing the (PE, Buffers) of two layers of a models and swap them. This conservative self-crossover preserves most of the learnt relationship between layer and resources and adds an exploration effect.

%% file: tables/action_values.tex
\begin{table}[t]
\centering
\caption{The level values of action pair.}
\includegraphics[width=1\linewidth]{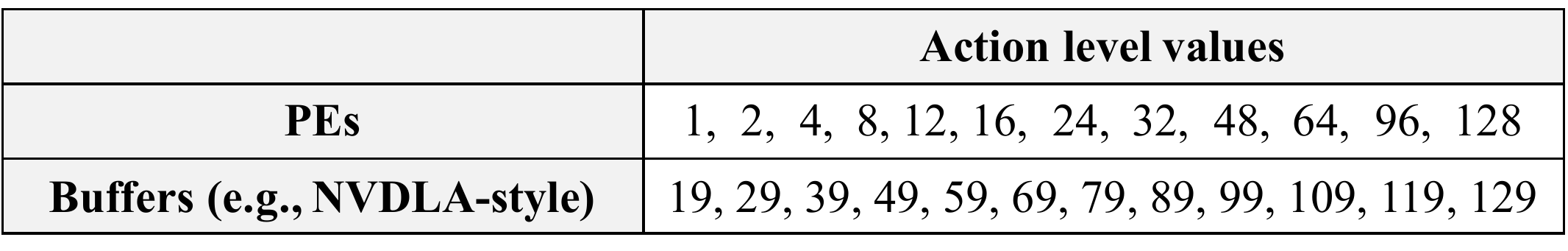}
\vspace{-0.4cm}
\label{table:action_value}
\end{table}

%% file: sections/04-Evaluation.tex
\vspace{-0.1cm}
\section{Evaluations}
\label{sec:experiments}
\vspace{-2mm}

\input{tables/table_platform}

\subsection{Methodology}

\subsubsection{\textbf{DNN Models}}
In our evaluations, we consider three
CNN models with different complexity: \mobilenet \cite{sandler2018mobilenetv2}, \mnasnet \cite{tan2019mnasnet}, and \resnet \cite{resnet}.
We also evaluate three GEMM-based ML models: GNMT ~\cite{gnmt} for machine translation, transformer~\cite{vaswani2017attention} for language understanding and NCF ~\cite{he2017neural} for collaborative filtering.

\subsubsection{\textbf{Accelerator Platforms}} We consider three different classes of platforms: Cloud server, IoT device and extremely small IoT, and, for comparison, an unconstrained platform as shown in \autoref{table:platform}.
We consider three dataflows: NVDLA-style \cite{nvdla} (-dla)  (parallelizing K and C dim.), Eyeriss-style \cite{chen2016eyeriss} (-eye) (parallelizing Y and R dim.), ShiDianNao-style \cite{du2015shidiannao} (-shi)(parallelizing Y and X dim).
We use L=12 levels of action values for PE and Buffers, where $(p_{n_{th}}, b_{k_{th}})$ represents assigning $n_{th}$-level of PEs and $k_{th}$-level of Buffers.



\input{figure/onelayer}
\input{tables/combined_dnn_gemm}

\input{tables/compare_GA}

\input{tables/compare_rl}

\subsubsection{\textbf{Baseline Optimization/Search Methods}}
We evaluate the following optimization methods as baselines.

\textbf{Grid search.} We enumerate through the design space with the stride of $s$ in the L=12 level, (e.g., $(p_{1_{th}}, b_{1_{th}})$, $(p_{1_{th}}, b_{(1+s)_{th}})$...). We set maximum epochs $Eps$. We emulate through the design space until the number of sampling points reached $Eps$.

\textbf{Random search}. We randomly sample $Eps$ design points and keep the best solutions as a result.


\textbf{GA~\cite{ga}.}
The baseline is a general GA algorithm, not the specially designed local fine-tuning one as described in \autoref{subsec:our_ga}. The GA is set with 100 population, and $\left \lceil \frac{Eps}{100} \right \rceil$ generations. The mutation rate and crossover rate is set as 0.05.

\textbf{Simulated Annealing \cite{SA}.}  The simulated annealing is implemented with temperature of 10 with step size of 1 and adopted to discrete integer space. 

\textbf{\bayes optimization \cite{bayes}.}  We set the algorithm to run for $Eps$ iterations, where $Eps$ points are sampled by the algorithm. We adopt it to discrete integer space. We set the number of the optimizer to 5 for the Gaussian process, since we empirically find this setting has better performance. 

\textbf{State-of-the-art RL algorithms.}
We consider state-of-the-art RL algorithms that are successful in many control problems. We consider both continuous and discrete methods. We compare with \textbf{A2C} \cite{a2c}, \textbf{ACTKR} (Actor Critic using Kronecker-Factored Trust Region) \cite{acktr}, and \textbf{PPO2} \cite{ppo2}. Both continuous and discrete versions of the three algorithms are experimented. Across all the experiments, we found the discrete version converge to better value. Hence, we will only show the result of the discrete version in the comparisons table. We also consider \textbf{DDPG} \cite{ddpg}, \textbf{SAC} \cite{sac}, and \textbf{TD3} \cite{td3} in continuous space. All comparisons run for $Eps$ epochs.

\subsubsection{\textbf{\system (Global)}}
We only consider the first-stage global search, Con'X (global), throughout the comparisons against baseline methods, for fairness. The second-stage fine-tuning can be added on top of the first-stage results. The benefit of the second-stage is explicitly discussed in \autoref{sec:eval_2stage}.


\subsection{Per-layer study for LS deployment}
We start by showing the HW performance of different action pairs $(p_{n_{th}}, b_{k_{th}})$ with 12 level of values each, which is Y-axis and X-axis in \autoref{fig:onelayer}. We sweep through $(p_{n_{th}}, b_{k_{th}})$ with exhaustive search and color it with their corresponding latency/energy value. Red indicates large latency/energy values while purple indicates small ones. For each layer, the contour is drastically different. Each layer require distinct action pairs $(p_{n_{th}}, b_{k_{th}})$ to reach optimal values (purple). The contour becomes a flat region when the PEs or Buffers are over-provisioned. Latency of layer-12 as an example, when PE is larger than the $9_{th}$ level and Buffer is larger than the $3_{rd}$ level, the latency remains the same because of over-provisioning. The two separate purple region in latency of Layer-34 indicates that there are two region of tiling size, which we map to the buffer, can optimize the latency. For Layer-23 (DWCONV), increasing the tile size of the mapping dimension (K) does not help because of the irrelevance of each output channel (K) in DWCONV. As for energy, larger number of PEs and Buffers can potentially decrease the energy because of shorter execution time as in layer-12 and layer-34. We can observe there are sweet spot for buffer size in Layer-23, where all the channel is mapped to one PE. At this end, increasing PE would not increase the energy, since extra PE will be idle. Also decreasing Buffers cause more times of fetching, which increase the energy consumption.

Con'X consistently finds the optimal action pair for each layer, and its solution is as-good or better (fewer PEs and buffers for same latency or energy) than the baseline methods and two common heuristics.
\autoref{fig:onelayer} also shows that there is no action pair that suits all the layers.
Thus, for a LS scenario, a designer can use Con'X to find optimal configurations for each layer, and then pick the one that provides optimum values across most layers.

\subsection{LP Deployment}
Next, we consider LP deployment (i.e., all layers of the model mapped on the accelerator) with platform constraints. For all the comparisons, we compare the algorithm performance by comparing their best solutions after $Eps=5,000$ epochs.

\subsubsection{\textbf{Converged solutions across DNNs, Dataflows, and Platforms}}
We ran baseline optimization methods and RL algorithms for a suite of DNNs (CNN- and GEMM-based) with varying dataflow styles and platform constraints. The objective is set to minimize the latency of the entire model. Therefore the lower the reached value, the better the solution is.
In the interest of space, we show the results with the best performing baselines, GA and PPO2, in \autoref{table:combined}. GA can reach good optimized value when the constraint is loose (cloud), but it fails in some tight constraint cases (IoT, IoTx). Both PPO2 and Con'X(global) can find solutions in any type of constraint. Across all the experiments, Con'X(global) finds the solution with the same or better performance than PPO2 and GA.

\subsubsection{\textbf{Deep-dive with optimization methods}}
\autoref{table:compare_ga} compares the 
solutions attained by various optimization methods and Con'X(global) for \mobilenet under four platform constraints for a NVDLA-style accelerator.
The objective is set to minimize the latency or energy of the entire model. 
Random, SA, and GA fail to come up with a feasible solution when faced with tight constraint (IoT). Also, \bayes optimization fails in extreme tight constraint (IoTx). Con'X(global) successfully learns the constraint behavior and optimizes the objective together. Con'X(global) generates the most optimized design points with 86\% lower latency and 70\% lower energy, on average across baselines.

\input{figure/critic_network}

\subsubsection{\textbf{Deep-dive with RL algorithms}}
\label{subsec:compare_rl}
We compare Con'X(global) with other state-of-the-art RL algorithms in the same setting as the previous experiment, as shown in \autoref{table:compare_rl}. All the RL agents are able to find feasible solutions in all situations. Considering the complexity of the algorithm, DDPG \cite{ddpg}, SAC \cite{sac}, and TD3 \cite{td3} generally consume more search time and memory overhead. Across all comparisons, we find Con'X(global) and PPO2 \cite{ppo2} reach better objective value. Con'X(global) converges to the optimized value 4.7 to 24 times faster than alternate RL algorithms.

\textbf{Analysis of critic networks.}
In many advanced RL algorithms such as A2C~\cite{a2c}, ACTR~\cite{acktr}, PPO2~\cite{ppo2}, DDPG~\cite{ddpg}, SAC~\cite{sac} and TD3~\cite{td3}, critic networks are used
to approximate the underlying value functions, which in turn train the policy network.
The REINFORCE-based used in \system{}, on the other hand, only has an actor network that learns directly from the reward.
As \autoref{table:compare_rl} shows, we found that REINFORCE~\cite{reinforce} in Con'X(global) converges to better solutions than all the actor-critic RL algorithms.
Our intuition is that this is because the function of the HW performance of the accelerator are too discrete and irregular for a critic neural network to learn well, and this in turn
adversely affects the learning of the policy networks.
To verify this intuition, we extract the critic network from the implemented alternate RL algorithms \cite{ppo2, ddpg, td3, sac, a2c, acktr} and conduct a standalone experiment to test its ability to approximate the underlying value function. The task is to take the ``state values" as input and predict the corresponding reward of that state. We use per-layer latency of \mobilenet as reward.
We use mean square error and gradient decent to train the network. We show the root mean square error (RMSE) when training with different size of data, as shown in \autoref{fig:critic_network}. 260,000 is the maximum possible data points critic network can experience under the RL tasks of $Eps=5,000$ with \mobilenet. We can observe that the training and testing loss is hard to converge to a feasible value (the best RMSE is 5.3e+4, which means the predicted latency (reward) by critic network is in average 5.3e+4 cycles difference to the ground-truth ones) which means the critic network did not learn reward value well. This could potentially misguide the policy network.

\input{figure/sample_efficiency_fig}
\subsubsection{\textbf{Sample efficiency and convergence}}
In the experiments against baseline optimization methods and other RL algorithms, we found Con'X(global) has the fastest convergence rate. We show two convergence traces as examples in \autoref{fig:sample_efficiency} for \mobilenet. With rapid convergence, our method heads toward the objective with more sample efficiency. On the contrary, the exhaustive search needs to enumerate and search through $L^{2N}$, $12^{104}$=$O(10^{112})$, data points for the search space of 52-layer \mobilenet with two actions per layer and 12 level of values per action, which is near impossible to finish.

\input{tables/mix_dataflow}
\input{figure/mix_dataflow_barchart}
\subsection{Dataflow-HW co-automation}
We extend \system{} to co-automate the per-layer dataflow style decision. Rather than manually picking one of the dataflow style, we let the agent make this decision. To take one step further, we let the agent do fine-grained per-layer dataflow style decision, which we termed as MIX-strategy. The agent now makes three decisions per-layer: PEs, Buffers, and dataflow style.  We found Con'X-MIX can not only pick the best dataflow-style for a model but also take advantage of MIX-strategy to pick different dataflow-style in different layers, as shown in \autoref{fig:mix_dataflow_barchart} for \mobilenet.In general, if there are no HW resource constraints, system will favor eye/shi at early layers (larger activations), which parallelize along activations dimensions, and favor dla at late layers (larger K/C), which parallelizes along channel dimensions (K/C) in CNN-based networks. However, when considering HW constraint, it becomes a compound decision trading-off among PE, Buffers, dataflow-style, and area. From the experiment listed in \autoref{table:mix_dataflow}, we can observe that in a more relaxed constraint, dla performs better than the other two since most layers in CNN-based networks have large K/C dim. However, in a tighter constraint, the parallelization ability of dla will be restricted; Eye/shi, which parallelize activations dim (whose values shrink layer-by-layer quickly in most CNNs) become more efficient choices. This observation can also explain the fact that system chooses eye/dla for some of the later layers in \autoref{fig:mix_dataflow_barchart}. From the experiments listed in \autoref{table:mix_dataflow}, Con'X-MIX further improves the optimization results by 4\% to 69\% comparing to the best-performing Con'X-dla/shi/eye. 

\input{tables/finetune}
\input{figure/rl_finetuning}
\subsection{Benefit of Two-stage Optimization}
\label{sec:eval_2stage}
In the above comparing with baseline experiments, we did not use local fine-tuning for fairness. We now show its effectiveness. We use local GA of 20 populations and run for 2,000 generations. We use local crossover rate of 0.2, local mutation rate of 0.05, and local mutation step of 4 .

\subsubsection{\textbf{The effect of fine-tuning}}
We show the two stage optimization results in \autoref{table:finetune}. We show one of the trace of reached-value along epochs in \autoref{fig:rl_finetuning}, which is the first row in \autoref{table:finetune} and the third row in \autoref{table:compare_ga}. In this case, pure GA cannot find valid solution because of the tight constraint (IoT).
The first-stage global search of Con'X learns to generate a valid solution first, whose value is recorded as initial valid value in \autoref{table:finetune}. Then, Con'X starts to optimize the value while conforming to constraint and reached an optimized point. The first stage improves the values from 56\% to 99\% compared to the initial valid values. Then, local fine-tuning using GA to further optimizes the solutions, and they improves by another 7\% to 93\% than the output of the first stage, which are 66\% to 99\% improvement over the initial value.

\input{figure/sol_reasoning}

\subsubsection{\textbf{Analysis of Design-Points found by \system}}
In \autoref{fig:sol_reasoning}, at the top, we show how \system allocates area to different components (total PE, total buffers, and per-layer) for \mobilenet and \resnet in an experiment with total area constrained.
The per-layer assignment is highly heterogeneous, which 
can be seen by the per-layer PE and Buffer assignment shown at the bottom of \autoref{fig:sol_reasoning}. 
In particular, in \mobilenet, we observe that the DWCONV layers
are assigned less resources on both PEs and Buffers. This could be because they require less computation and we are limited by the platform area constraint, which makes the agent reduce the assigned resources.
In \resnet, we find that the agent assigns more Buffers to the layers that have the larger number of input/output channel size (e.g., layers 37,43, and 47).


\input{tables/scenario_c}

\input{tables/policy_network}

\subsection{LP deployment at compile time}
\system can also be used for LP deployment at compile time. One common use-case is for FPGA-based accelerator design. As is common for FPGAs, we impose the maximum number of PEs and Buffers as constraint (which would depend on the specific FPGA board). 
We consider both cloud and edge FPGAs as constraints. 
The baseline is configured with uniform number of PE and Buffers for each layer with NVDLA-style dataflow. In \autoref{table:scenario_c}, we show that Con'X(global)-dla performs better than baseline-dla. Then we show that local-finetuning in Con'X(global)-dla can further improves the value by 7\% to 36\%. Finally, we show the two stage results of \system-MIX, where the final reached value is 50\% to 72\% better than baseline-dla.

\subsection{Policy Network Exploration}
\label{subsec:policy_network}
We show our design decision process of the policy network. First is the action levels, L, where we pick L=12, in the experiments. By decreasing L, we decrease the complexity of the problem but worsen the granularity, and vice versa. As shown in \autoref{table:policy_network}, L=12 is the sweet spot we found. We also experimented with different type of policy networks: MLP-based and RNN-based, as \autoref{table:policy_network} shows. We found RNN-based networks converging to better results, which may be owing to the fact that RNN is taking advantage of remembering the consumed constraint of previous layers.

\subsection{Summary}
We summarize some key results here.
For global search, we observe that RLs can explore an extremely large design space more effectively and efficiently compared to the baseline optimization methods. 
Next, we find that REINFORCE, which does not rely on value network, can converge faster and reach similar or better results than alternate RL methods in the discrete and irregular HW performance exploration problem. Next, we demonstrate that our formulation of REINFORCE-based (Con'X(global)) can not only explore the HW configuration but also effectively explore the dataflow-style decision simultaneously and further optimize the results by 4\% to 69\%. Finally, after a coarse-grained solution is found, we show that using a specialized GA for fine-tuning the result locally can optimize the result by another 7\% to 93\%.


%% file: tables/table_platform.tex
\begin{table}[t]
\centering
\caption{Platform constraint settings.}
\includegraphics[width=1\linewidth]{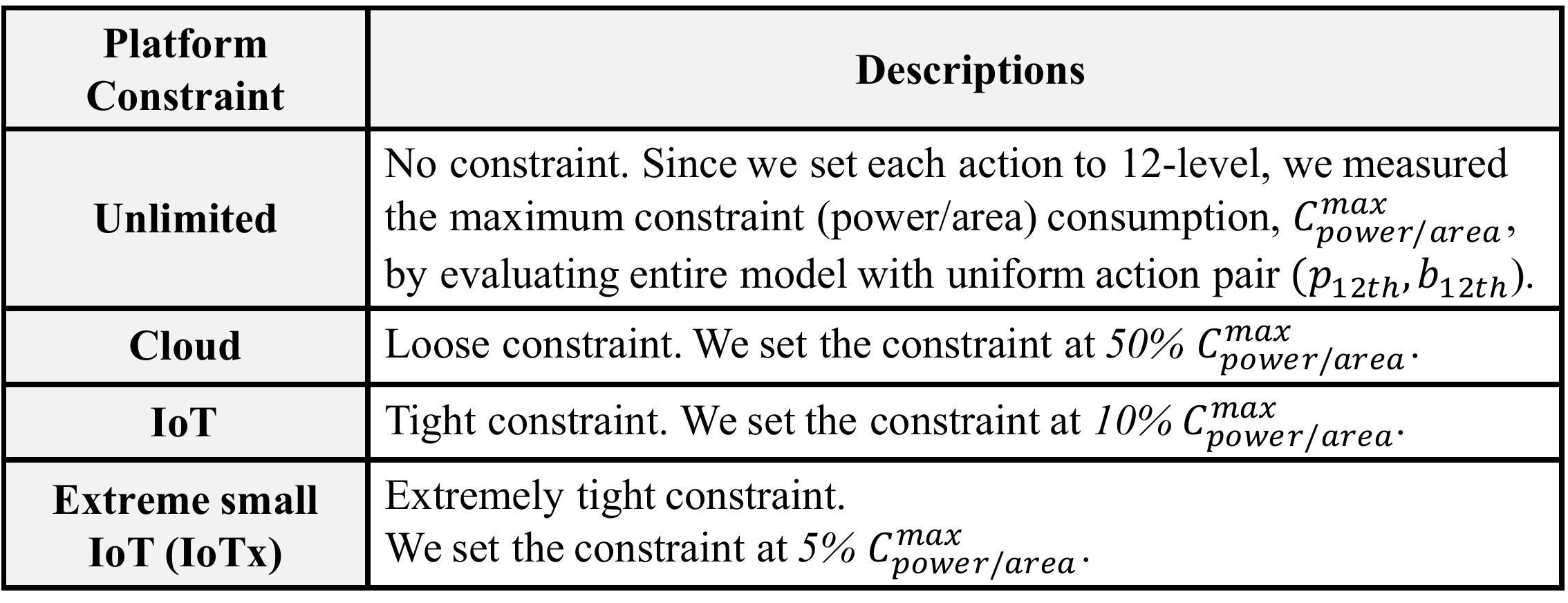}
\vspace{-0.6cm}
\label{table:platform}
\end{table}

%% file: figure/onelayer.tex
\begin{figure*}
\begin{center}
\includegraphics[width=0.8\linewidth]{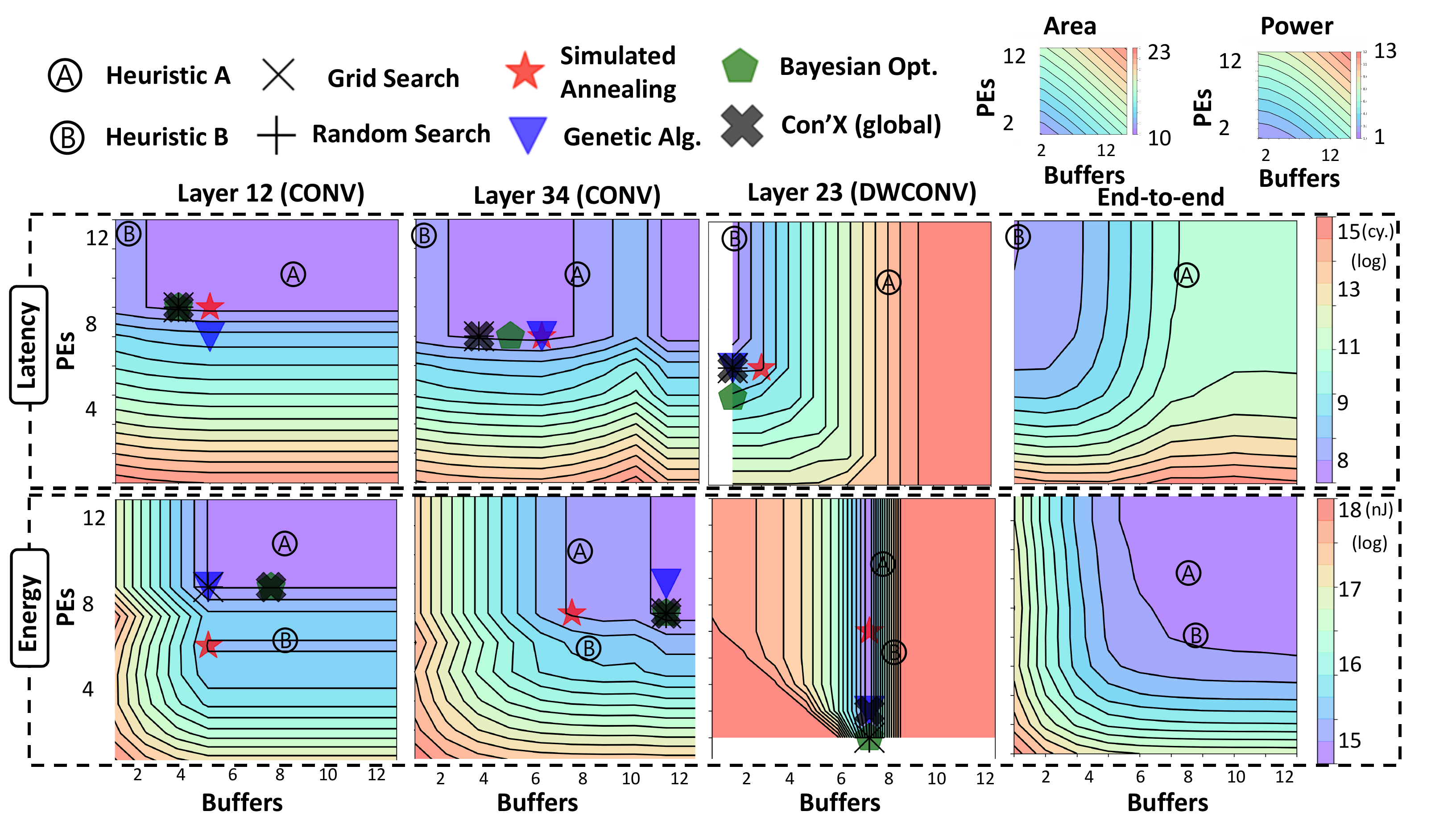}
\end{center}
\vspace{-0.2cm}
\caption{Searching for per-layer PEs/Buffers configurations to optimize latency/energy with different techniques. Purple indicates lower (better) and red indicates higher (worse) latency/energy. \textbf{Heuristic A}:  Determine the PEs/Buffers with the most compute-intensive layer (Layer-38) and apply the same configuration for all the layers. \textbf{Heuristic B}: Determine the PEs/Buffers by the configuration that optimizes end-to-end whole model latency/energy.}

\label{fig:onelayer}
\end{figure*}

%% file: tables/combined_dnn_gemm.tex
\begin{table}[t]
\centering
\caption{Converged solution of LP deployment.}
\includegraphics[width=1\linewidth]{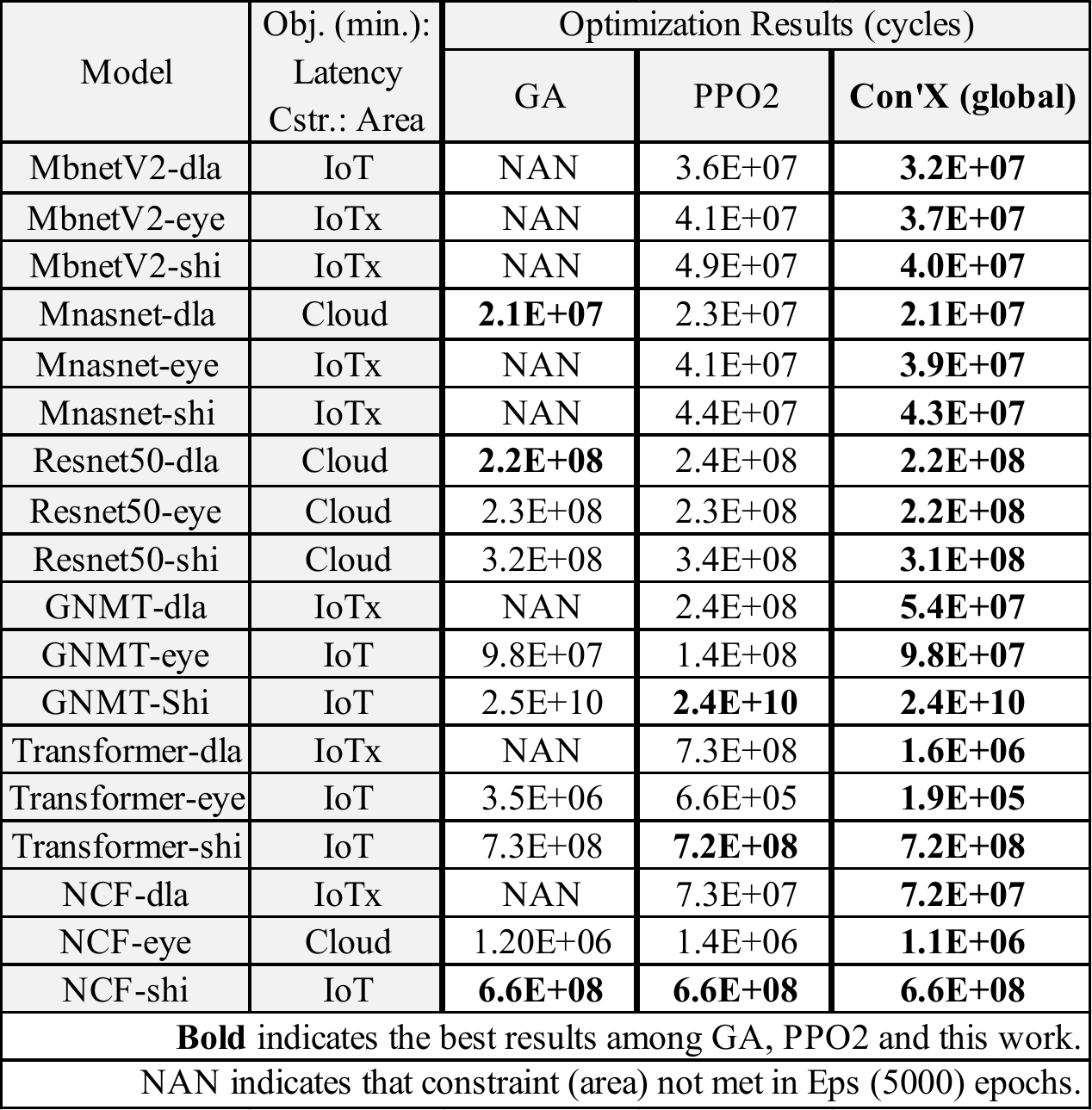}
\vspace{-0.4cm}
\label{table:combined}
\end{table}

%% file: tables/compare_GA.tex
\begin{table}[t]
\centering
\caption{Converged solutions after 5000 epochs for various optimization methods across 
four platforms with different constraints. DNN=\mobilenet, Dataflow=NVDLA-style, Deployment=LP}
\includegraphics[width=1\linewidth]{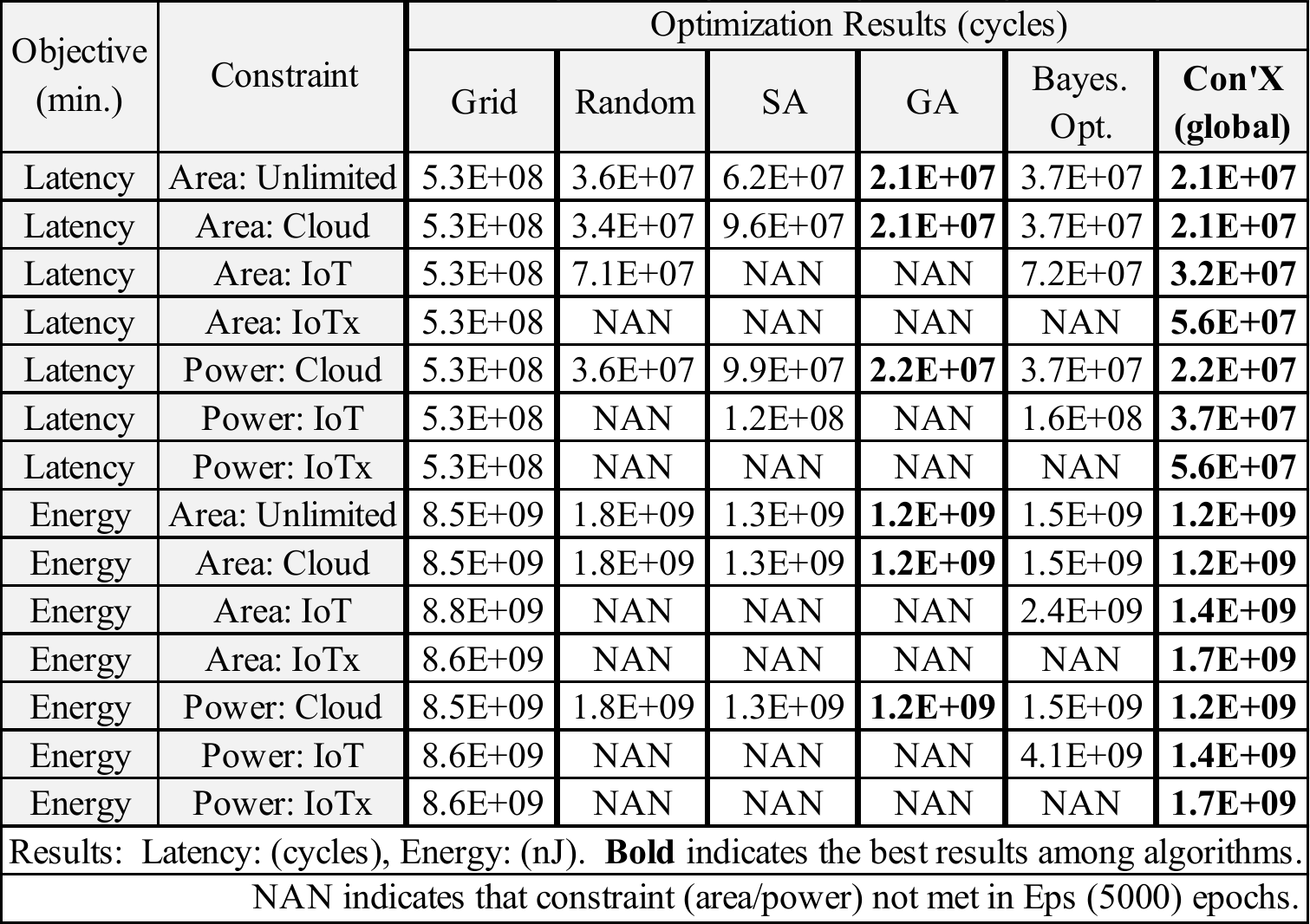}
\vspace{-0.2cm}
\label{table:compare_ga}
\end{table}

%% file: tables/compare_rl.tex



\begin{table*}[t]
\centering
\caption{Comparison of search-time and converged solutions across state-of-the-art RL techniques.} 
\includegraphics[width=1\linewidth]{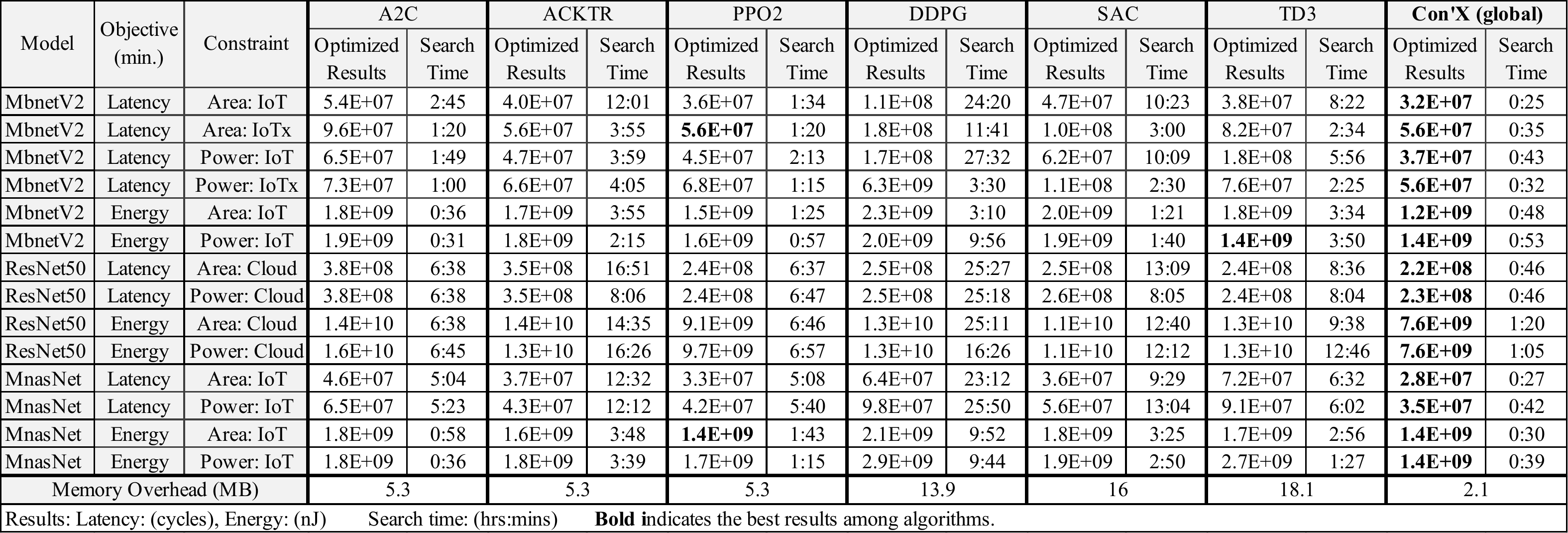}
\vspace{-0.6cm}
\label{table:compare_rl}
\end{table*}

%% file: figure/critic_network.tex
\begin{figure}[t]
\begin{center}
\includegraphics[width=1\linewidth]{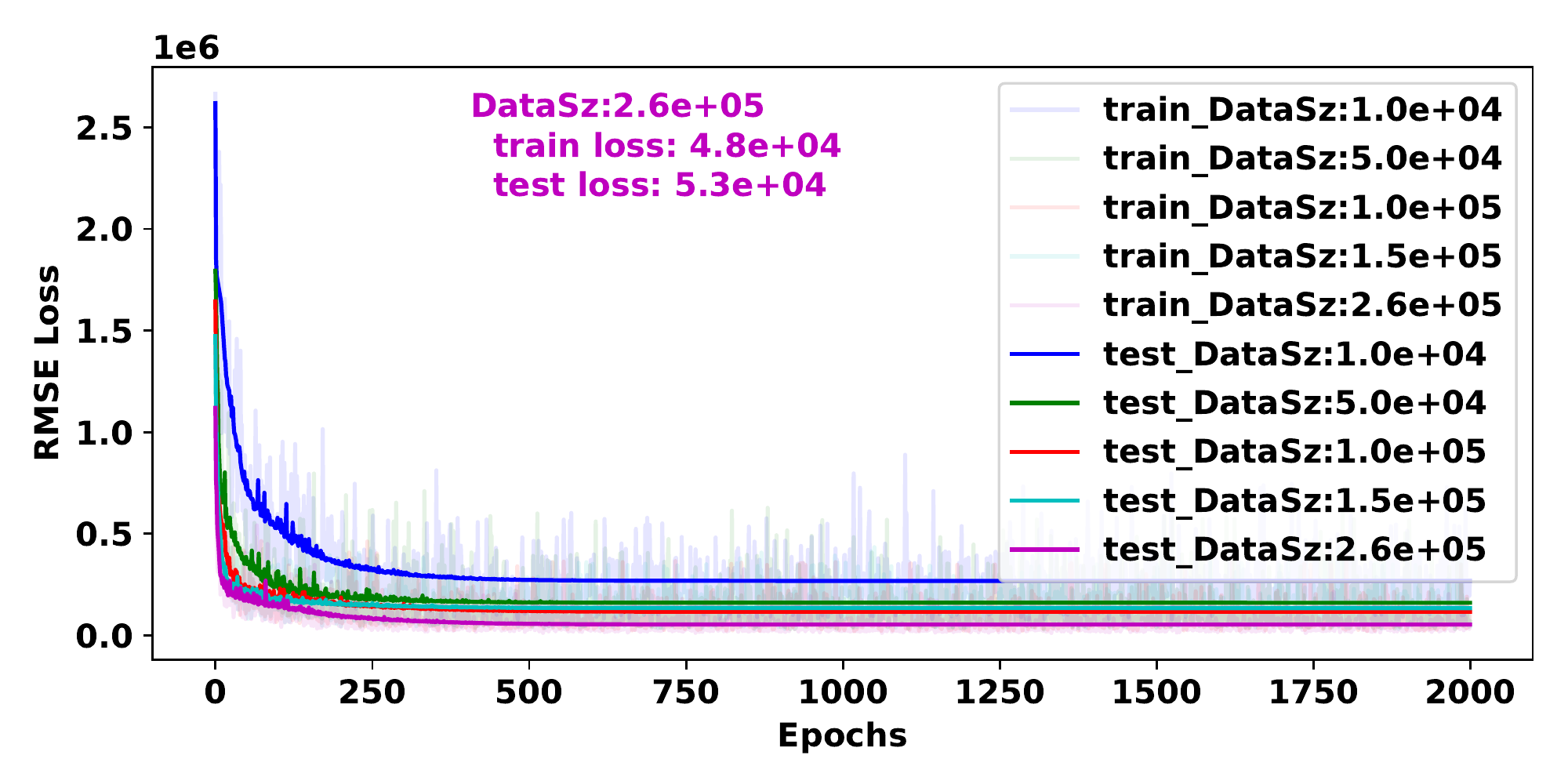}
\end{center}
\vspace{-0.40cm}
\caption{The learning curve of the critic network.}
\vspace{-0.1cm}
\label{fig:critic_network}
\end{figure}

%% file: figure/sample_efficiency_fig.tex
\begin{figure}
\begin{center}
\includegraphics[width=1\linewidth]{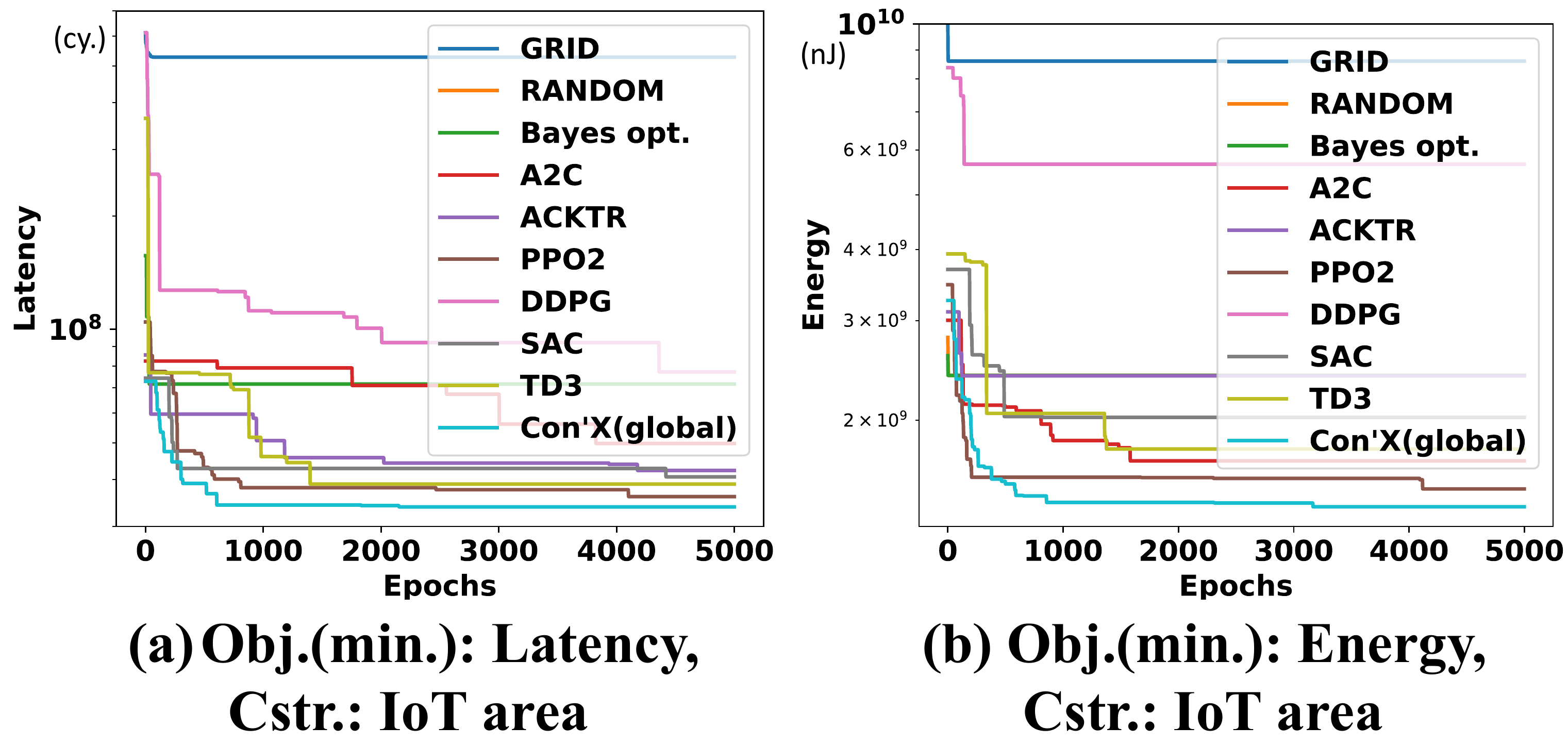}
\end{center}

\caption{The fast convergence and sample efficiency of Con'X (global).}

\label{fig:sample_efficiency}
\end{figure}

%% file: tables/mix_dataflow.tex
\begin{table}[t]
\centering
\caption{Dataflow and Hardware co-automation.}
\includegraphics[width=1\linewidth]{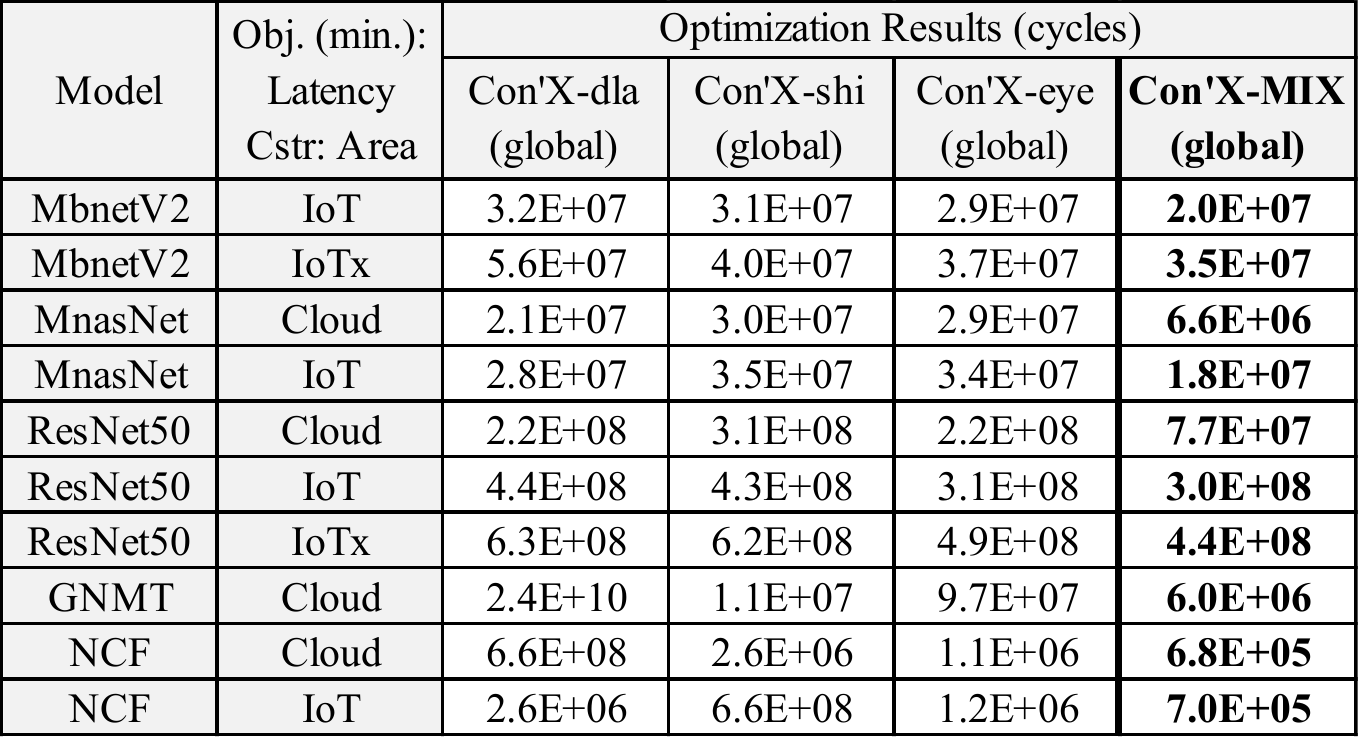}
\label{table:mix_dataflow}
\vspace{-0.4cm}
\end{table}
\vspace{-2mm}

%% file: figure/mix_dataflow_barchart.tex
\begin{figure*}[t]
\begin{center}
\includegraphics[width=1\linewidth]{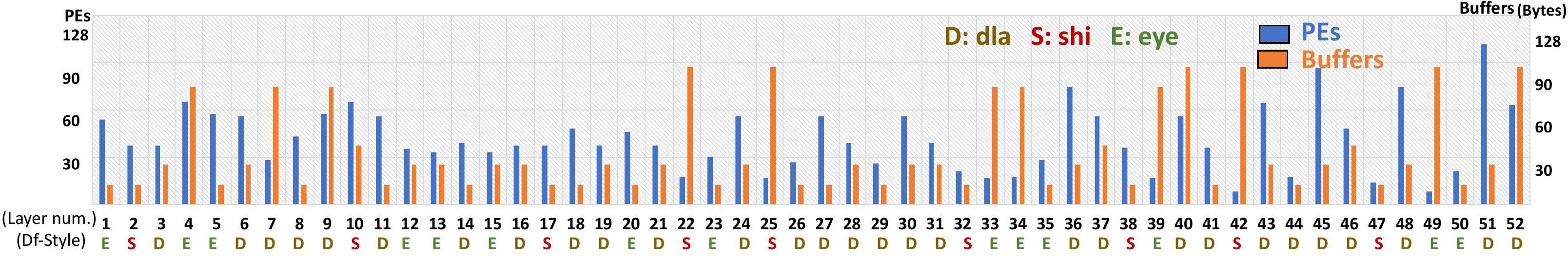}
\end{center}

\caption{Dataflow-HW co-automation with \system for \mobilenet. (Obj.(min.):Latency,Cstr:IoT area).}
\vspace{-0.1cm}
\label{fig:mix_dataflow_barchart}
\end{figure*}

%% file: tables/finetune.tex
\begin{table}[t]
\centering
\caption{Two-stage optimization of \system.}
\includegraphics[width=1\linewidth]{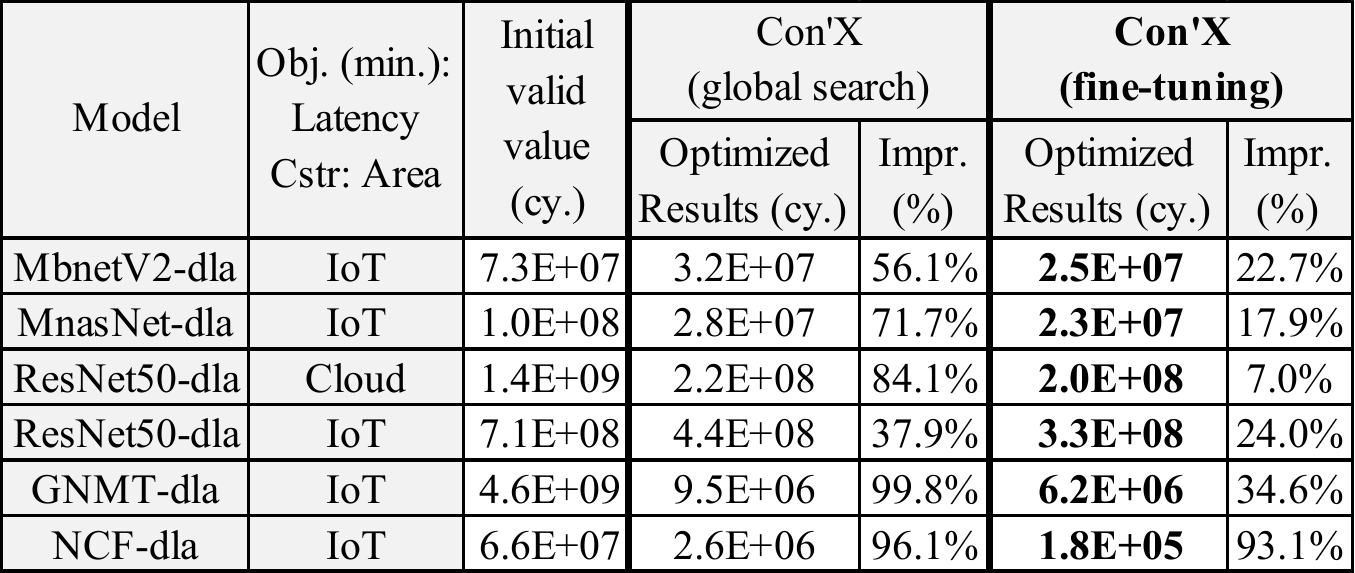}
\vspace{-0.2cm}
\label{table:finetune}
\end{table}

%% file: figure/rl_finetuning.tex
\begin{figure}
\begin{center}
\includegraphics[width=1\linewidth]{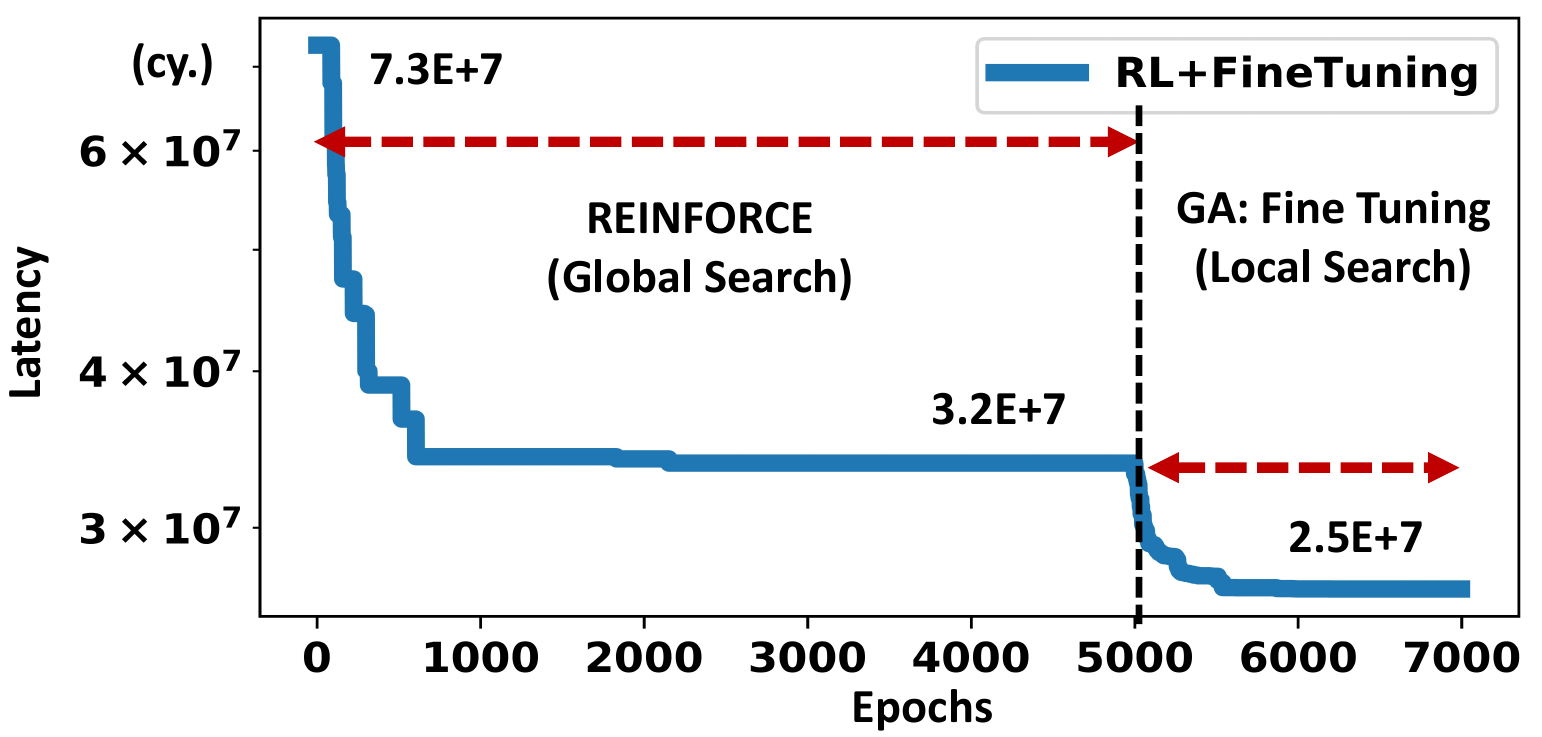}
\end{center}

\caption{ Overall latency as a function of epochs across two-stage optimization in \system{} (\mobilenet, Obj.(min.):Latency, Cstr.:IoT area). }

\label{fig:rl_finetuning}
\end{figure}

%% file: figure/sol_reasoning.tex
\begin{figure*}[t]
\begin{center}
\includegraphics[width=1\linewidth]{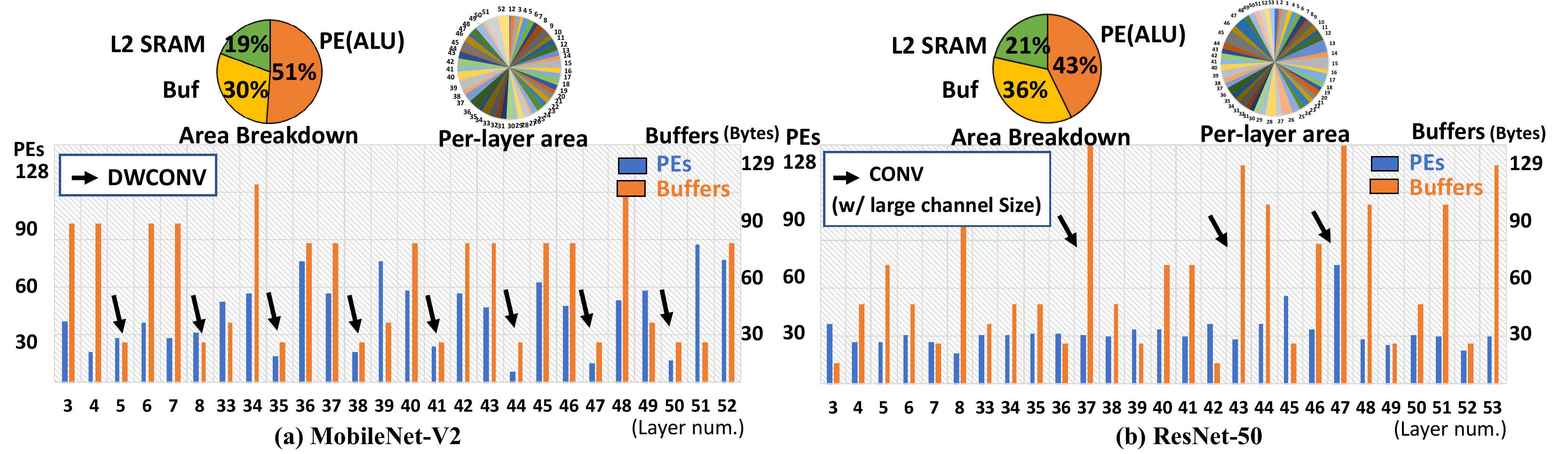}
\end{center}

\caption{The solution for (a) \mobilenet and (b) \resnet (Obj.(min.):Latency, Cstr:IoT area).}
\vspace{-0.1cm}
\label{fig:sol_reasoning}
\end{figure*}

%% file: tables/scenario_c.tex
\begin{table*}[h]
\centering
\caption{Resource assignments for LP deployment at compile time of \system.}
\includegraphics[width=1\linewidth]{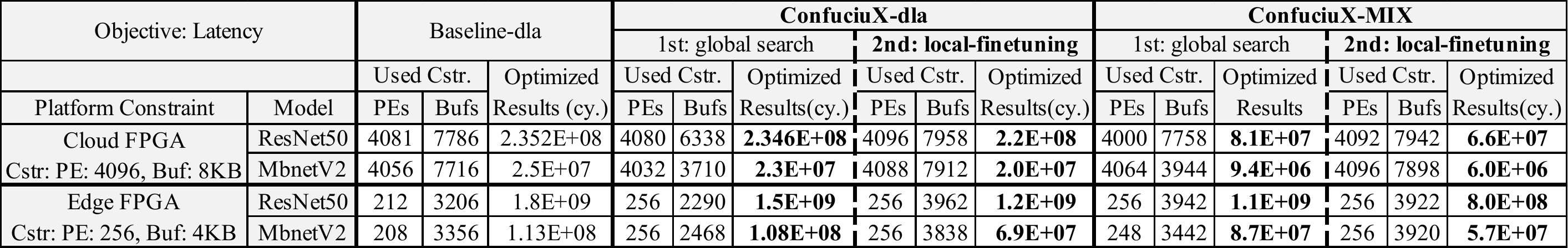}
\vspace{-0.4cm}
\label{table:scenario_c}
\end{table*}

%% file: tables/policy_network.tex
\begin{table}[t]
\centering
\caption{Different configurations of the policy network.}
\includegraphics[width=1\linewidth]{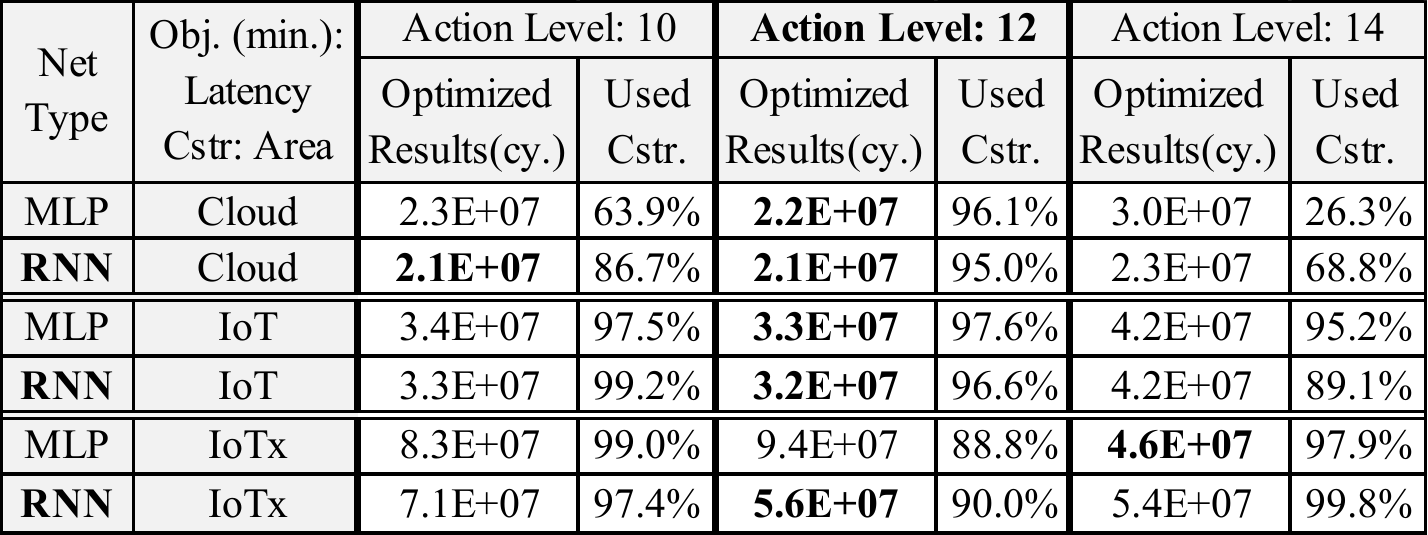}
\vspace{-0.2cm}
\label{table:policy_network}
\end{table}

%% file: sections/05-Relatedwork.tex
\section{Related Works}
\label{sec:relatedworks}

\textbf{Accelerator HW Design-Space Exploration.}
Fine-grained HW resource assignment has been studied extensively for LS deployment on FPGAs \cite{hao2019fpga,zhong2017design,motamedi2016design}. 
Whole-model LP deployment has been shown to be more efficient than LS deployments with uniform resource assignments for every layer~\cite{shen2017maximizing,li2016high,zhang2016energy,jiang2018heterogeneous,jiang2019xfer,zhang2018dnnbuilder,chen2019cloud,wei2018tgpa}. Many works have focused on allocating resources for convolution layers in a LP deployment within one FPGA \cite{shen2017maximizing, zhang2018dnnbuilder}, across multiple FPGAs \cite{jiang2018heterogeneous,li2016high,jiang2019xfer, wei2018tgpa,jiang2019achieving} or in cloud FPGA platforms \cite{chen2019cloud}. Some works have focused on HW DSE for ASIC accelerators\cite{santoro2018energy, minerva, aladdin} or templated systolic array structures~\cite{cong2018polysa, wei2017automated}. Some general frameworks execute the design space exploration at the architecture level, supporting both ASIC and FPGA \cite{yang2018dnn, maestro}. 
Yang et. al, \cite{yang2018dnn} further shows that the HW resource assignment dominates the performance of accelerator comparing to dataflow exploration.
For design space exploration, most of these prior
works employ grid/exhaustive search, while techniques for pruning the exploration spaces are manually developed. However, with myriads of DNN models being designed on a daily basis, it becomes harder to manually design and tune the policy for the newly constructed search space. 
In this work, we develop a ML-based method to automate the search process with high sample efficiency for both LP and LS scenario.

\textbf{ML-based methods for DNN compilation and mapping.}
ML methods have 
found value in mapping/compiling DNNs over hardware.
TensorComprehensions ~\cite{vasilache2018tensor} uses genetic algorithm,
AutoTVM ~\cite{chen2018tvm,autotvm} uses simulated annealing and boosted tree, Reagen et. al,~\cite{reagen2017case} uses \bayes optimization, RELEASE~\cite{ahn2019reinforcement} uses RL, ATLAS~\cite{whaley1998automatically} uses black box optimizations, some compiler design~\cite{novillo2014samplepgo, chang1991using} use profile-guided optimization to perform target-independent front-end compiler optimizations on DNNs or linear algebra computations. 
Some recent works use RL on HW/SW co-exploration to explore both DNN and its mapping over hardware \cite{jiang2019hardware,abdelfattah2020best,lu2019neural,yang2020co}.
The problem of mapping the DNN computation graph over multiple devices (CPU/GPU/TPU~\cite{tpu})
has also been explored through manual heuristics \cite{sutskever2014sequence,bahdanau2014neural,yonghui2016bridging} and RL \cite{mirhoseini2017device, spotlight,paliwal2020reinforced}.
In contrast to these works, 
this work looks at fine-grained design-time assignment of compute and memory within an accelerator.

\textbf{Dataflow style optimization.} Architecture design of ML accelerators include resource assignment and dataflow style design. Dataflow style is a scheduling and compiler optimization problem, which has been studied for decades for the generic platform such as CPU or GPU \cite{dmazerunner,abadi2016tensorflow,wei2017dlvm,ragan2013halide,kjolstad2017tensor,klockner2014loo,steuwer2017lift,lane2016deepx}, or for FPGA \cite{ma2017optimizing,cong_fpga}, while they apply grid search for reaching their objectives. For ML accelerators, some mainstream dataflow style are manually designed and proven to be efficient, becoming prominent or commercialized \cite{chen2016eyeriss, tpu, du2015shidiannao, nvdla}.
In this work, we focus on the resource assignment part of the accelerator design flow, and utilize some prominent dataflow styles \cite{chen2016eyeriss, tpu, du2015shidiannao, nvdla}.


%% file: sections/06-Conclusion.tex
\section{Conclusion}
\label{sec:conclusion}
While efficient DNN models and dataflow style are widely studied for ML accelerators, HW resource assignment is relatively unexplored. In this paper, we propose \system, an autonomous strategy to find out the optimized HW resource assignment for a given DNNs, a dataflow style and platform constraints. \system leverages RL for the global search, augmented with GA for fine-tuning. We quantitatively experiment on different models, platform constraints and dataflow styles. \system demonstrates the highest sample-efficiency compared to other optimization and RL methods.
This works shows the promise of leveraging ML within the DNN accelerator design workflow, with opportunities for future work across new ML algorithms for learning dataflow/hardware behavior, and DNN-dataflow-hardware co-design.




%% file: sections/07-Acknowledgment.tex
\section*{Acknowledgement}
This work was supported by NSF Award 1909900 and a Google Faculty Award.
We thank Hyoukjun Kwon for help
with MAESTRO setup and feedback on the writing.
We acknowledge Arun Ramamurthy, Siva Theja Maguluri and Ananda Samajdar for helpful technical discussions.
A special thanks to Cliff Young for motivating us to pursue this 
research direction.

%% file: main.bbl
\begin{thebibliography}{10}
\providecommand{\url}[1]{#1}
\csname url@samestyle\endcsname
\providecommand{\newblock}{\relax}
\providecommand{\bibinfo}[2]{#2}
\providecommand{\BIBentrySTDinterwordspacing}{\spaceskip=0pt\relax}
\providecommand{\BIBentryALTinterwordstretchfactor}{4}
\providecommand{\BIBentryALTinterwordspacing}{\spaceskip=\fontdimen2\font plus
\BIBentryALTinterwordstretchfactor\fontdimen3\font minus
  \fontdimen4\font\relax}
\providecommand{\BIBforeignlanguage}[2]{{%
\expandafter\ifx\csname l@#1\endcsname\relax
\typeout{** WARNING: IEEEtranS.bst: No hyphenation pattern has been}%
\typeout{** loaded for the language `#1'. Using the pattern for}%
\typeout{** the default language instead.}%
\else
\language=\csname l@#1\endcsname
\fi
#2}}
\providecommand{\BIBdecl}{\relax}
\BIBdecl

\bibitem{stars_and_bars}
``Stars and bars (combinatorics),''
  \url{https://en.wikipedia.org/wiki/Stars_and_bars_(combinatorics)}.

\bibitem{openai}
``Openai gym,'' \url{https://gym.openai.com/}, 2016.

\bibitem{nvdla}
``Nvdla deep learning accelerator,'' \url{http://nvdla.org}, 2017.

\bibitem{maestro_web}
``Maestro tool,'' \url{http://maestro.ece.gatech.edu/}, 2020.

\bibitem{abadi2016tensorflow}
M.~Abadi, P.~Barham, J.~Chen, Z.~Chen, A.~Davis, J.~Dean, M.~Devin,
  S.~Ghemawat, G.~Irving, M.~Isard \emph{et~al.}, ``Tensorflow: A system for
  large-scale machine learning,'' in \emph{12th $\{$USENIX$\}$ Symposium on
  Operating Systems Design and Implementation ($\{$OSDI$\}$ 16)}, 2016, pp.
  265--283.

\bibitem{abdelfattah2020best}
M.~S. Abdelfattah, {\L}.~Dudziak, T.~Chau, R.~Lee, H.~Kim, and N.~D. Lane,
  ``Best of both worlds: Automl codesign of a cnn and its hardware
  accelerator,'' \emph{arXiv preprint arXiv:2002.05022}, 2020.

\bibitem{ahn2019reinforcement}
B.~H. Ahn, P.~Pilligundla, and H.~Esmaeilzadeh, ``Reinforcement learning and
  adaptive sampling for optimized dnn compilation,'' \emph{arXiv preprint
  arXiv:1905.12799}, 2019.

\bibitem{bahdanau2014neural}
D.~Bahdanau, K.~Cho, and Y.~Bengio, ``Neural machine translation by jointly
  learning to align and translate,'' \emph{arXiv preprint arXiv:1409.0473},
  2014.

\bibitem{bang201714}
S.~Bang, J.~Wang, Z.~Li, C.~Gao, Y.~Kim, Q.~Dong, Y.-P. Chen, L.~Fick, X.~Sun,
  R.~Dreslinski \emph{et~al.}, ``14.7 a 288$\mu$w programmable deep-learning
  processor with 270kb on-chip weight storage using non-uniform memory
  hierarchy for mobile intelligence,'' in \emph{2017 IEEE International
  Solid-State Circuits Conference (ISSCC)}.\hskip 1em plus 0.5em minus
  0.4em\relax IEEE, 2017, pp. 250--251.

\bibitem{bergstra2012random}
J.~Bergstra and Y.~Bengio, ``Random search for hyper-parameter optimization,''
  \emph{Journal of machine learning research}, vol.~13, no. Feb, pp. 281--305,
  2012.

\bibitem{chakradhar2010dynamically}
S.~Chakradhar, M.~Sankaradas, V.~Jakkula, and S.~Cadambi, ``A dynamically
  configurable coprocessor for convolutional neural networks,'' in
  \emph{Proceedings of the 37th annual international symposium on Computer
  architecture}, 2010, pp. 247--257.

\bibitem{chang1991using}
P.~P. Chang, S.~A. Mahlke, and W.-M.~W. Hwu, ``Using profile information to
  assist classic code optimizations,'' \emph{Software: Practice and
  Experience}, vol.~21, no.~12, pp. 1301--1321, 1991.

\bibitem{chen2018tvm}
T.~Chen, T.~Moreau, Z.~Jiang, L.~Zheng, E.~Yan, H.~Shen, M.~Cowan, L.~Wang,
  Y.~Hu, L.~Ceze \emph{et~al.}, ``$\{$TVM$\}$: An automated end-to-end
  optimizing compiler for deep learning,'' in \emph{13th $\{$USENIX$\}$
  Symposium on Operating Systems Design and Implementation ($\{$OSDI$\}$ 18)},
  2018, pp. 578--594.

\bibitem{autotvm}
T.~Chen, L.~Zheng, E.~Yan, Z.~Jiang, T.~Moreau, L.~Ceze, C.~Guestrin, and
  A.~Krishnamurthy, ``Learning to optimize tensor programs,'' in \emph{Advances
  in Neural Information Processing Systems}, 2018, pp. 3389--3400.

\bibitem{chen2019cloud}
Y.~Chen, J.~He, X.~Zhang, C.~Hao, and D.~Chen, ``Cloud-dnn: An open framework
  for mapping dnn models to cloud fpgas,'' in \emph{Proceedings of the 2019
  ACM/SIGDA International Symposium on Field-Programmable Gate Arrays}, 2019,
  pp. 73--82.

\bibitem{chen2016eyeriss}
Y.-H. Chen \emph{et~al.}, ``Eyeriss: An energy-efficient reconfigurable
  accelerator for deep convolutional neural networks,'' \emph{JSSC}, vol.~52,
  no.~1, pp. 127--138, 2016.

\bibitem{eyeriss_isscc}
{Chen, Yu-Hsin and Krishna, Tushar and Emer, Joel and Sze, Vivienne},
  ``{Eyeriss: An Energy-Efficient Reconfigurable Accelerator for Deep
  Convolutional Neural Networks},'' in \emph{{IEEE International Solid-State
  Circuits Conference, ISSCC 2016, Digest of Technical Papers}}, {2016}, pp.
  {262--263}.

\bibitem{cong2018polysa}
J.~Cong and J.~Wang, ``Polysa: polyhedral-based systolic array
  auto-compilation,'' in \emph{2018 IEEE/ACM International Conference on
  Computer-Aided Design (ICCAD)}.\hskip 1em plus 0.5em minus 0.4em\relax IEEE,
  2018, pp. 1--8.

\bibitem{dmazerunner}
S.~Dave, Y.~Kim, S.~Avancha, K.~Lee, and A.~Shrivastava, ``Dmazerunner:
  Executing perfectly nested loops on dataflow accelerators,'' \emph{ACM
  Transactions on Embedded Computing Systems (TECS)}, vol.~18, no.~5s, pp.
  1--27, 2019.

\bibitem{du2015shidiannao}
Z.~Du, R.~Fasthuber, T.~Chen, P.~Ienne, L.~Li, T.~Luo, X.~Feng, Y.~Chen, and
  O.~Temam, ``Shidiannao: Shifting vision processing closer to the sensor,'' in
  \emph{International Symposium on Computer Architecture (ISCA)}, 2015.

\bibitem{farabet2010hardware}
C.~Farabet, B.~Martini, P.~Akselrod, S.~Talay, Y.~LeCun, and E.~Culurciello,
  ``Hardware accelerated convolutional neural networks for synthetic vision
  systems,'' in \emph{Proceedings of 2010 IEEE International Symposium on
  Circuits and Systems}.\hskip 1em plus 0.5em minus 0.4em\relax IEEE, 2010, pp.
  257--260.

\bibitem{farabet2009cnp}
C.~Farabet, C.~Poulet, J.~Y. Han, and Y.~LeCun, ``Cnp: An fpga-based processor
  for convolutional networks,'' in \emph{2009 International Conference on Field
  Programmable Logic and Applications}.\hskip 1em plus 0.5em minus 0.4em\relax
  IEEE, 2009, pp. 32--37.

\bibitem{td3}
S.~Fujimoto, H.~Van~Hoof, and D.~Meger, ``Addressing function approximation
  error in actor-critic methods,'' \emph{arXiv preprint arXiv:1802.09477},
  2018.

\bibitem{spotlight}
\BIBentryALTinterwordspacing
Y.~Gao, L.~Chen, and B.~Li, ``Spotlight: Optimizing device placement for
  training deep neural networks,'' in \emph{Proceedings of the 35th
  International Conference on Machine Learning}, ser. Proceedings of Machine
  Learning Research, J.~Dy and A.~Krause, Eds., vol.~80.\hskip 1em plus 0.5em
  minus 0.4em\relax Stockholmsmässan, Stockholm Sweden: PMLR, 10--15 Jul 2018,
  pp. 1676--1684. [Online]. Available:
  \url{http://proceedings.mlr.press/v80/gao18a.html}
\BIBentrySTDinterwordspacing

\bibitem{sac}
T.~Haarnoja, A.~Zhou, P.~Abbeel, and S.~Levine, ``Soft actor-critic: Off-policy
  maximum entropy deep reinforcement learning with a stochastic actor,''
  \emph{arXiv preprint arXiv:1801.01290}, 2018.

\bibitem{hao2019fpga}
C.~Hao, X.~Zhang, Y.~Li, S.~Huang, J.~Xiong, K.~Rupnow, W.-m. Hwu, and D.~Chen,
  ``Fpga/dnn co-design: An efficient design methodology for 1ot intelligence on
  the edge,'' in \emph{2019 56th ACM/IEEE Design Automation Conference
  (DAC)}.\hskip 1em plus 0.5em minus 0.4em\relax IEEE, 2019, pp. 1--6.

\bibitem{resnet}
K.~He, X.~Zhang, S.~Ren, and J.~Sun, ``Deep residual learning for image
  recognition,'' in \emph{Proceedings of the IEEE conference on computer vision
  and pattern recognition}, 2016, pp. 770--778.

\bibitem{he2017neural}
X.~He, L.~Liao, H.~Zhang, L.~Nie, X.~Hu, and T.-S. Chua, ``Neural collaborative
  filtering,'' in \emph{Proceedings of the 26th international conference on
  world wide web}, 2017, pp. 173--182.

\bibitem{ga}
J.~H. Holland, ``Genetic algorithms,'' \emph{Scientific american}, vol. 267,
  no.~1, pp. 66--73, 1992.

\bibitem{jiang2019achieving}
W.~Jiang, E.~H.-M. Sha, X.~Zhang, L.~Yang, Q.~Zhuge, Y.~Shi, and J.~Hu,
  ``Achieving super-linear speedup across multi-fpga for real-time dnn
  inference,'' \emph{ACM Transactions on Embedded Computing Systems (TECS)},
  vol.~18, no.~5s, pp. 1--23, 2019.

\bibitem{jiang2018heterogeneous}
W.~Jiang, E.~H.-M. Sha, Q.~Zhuge, L.~Yang, X.~Chen, and J.~Hu, ``Heterogeneous
  fpga-based cost-optimal design for timing-constrained cnns,'' \emph{IEEE
  Transactions on Computer-Aided Design of Integrated Circuits and Systems},
  vol.~37, no.~11, pp. 2542--2554, 2018.

\bibitem{jiang2019hardware}
W.~Jiang, L.~Yang, E.~Sha, Q.~Zhuge, S.~Gu, Y.~Shi, and J.~Hu,
  ``Hardware/software co-exploration of neural architectures,'' \emph{arXiv
  preprint arXiv:1907.04650}, 2019.

\bibitem{jiang2019xfer}
W.~Jiang, X.~Zhang, E.~H.-M. Sha, Q.~Zhuge, L.~Yang, Y.~Shi, and J.~Hu, ``Xfer:
  A novel design to achieve super-linear performance on multiple fpgas for
  real-time ai,'' in \emph{Proceedings of the 2019 ACM/SIGDA International
  Symposium on Field-Programmable Gate Arrays}, 2019, pp. 305--305.

\bibitem{tpu}
N.~P. Jouppi, C.~Young, N.~Patil, D.~Patterson, G.~Agrawal, R.~Bajwa, S.~Bates,
  S.~Bhatia, N.~Boden, A.~Borchers \emph{et~al.}, ``In-datacenter performance
  analysis of a tensor processing unit,'' in \emph{International Symposium on
  Computer Architecture (ISCA)}.\hskip 1em plus 0.5em minus 0.4em\relax IEEE,
  2017, pp. 1--12.

\bibitem{SA}
S.~Kirkpatrick, C.~D. Gelatt, and M.~P. Vecchi, ``Optimization by simulated
  annealing,'' \emph{science}, vol. 220, no. 4598, pp. 671--680, 1983.

\bibitem{kjolstad2017tensor}
F.~Kjolstad, S.~Kamil, S.~Chou, D.~Lugato, and S.~Amarasinghe, ``The tensor
  algebra compiler,'' \emph{Proceedings of the ACM on Programming Languages},
  vol.~1, no. OOPSLA, pp. 1--29, 2017.

\bibitem{klockner2014loo}
A.~Kl{\"o}ckner, ``Loo. py: transformation-based code generation for gpus and
  cpus,'' in \emph{Proceedings of ACM SIGPLAN International Workshop on
  Libraries, Languages, and Compilers for Array Programming}, 2014, pp. 82--87.

\bibitem{maestro}
H.~Kwon, P.~Chatarasi, M.~Pellauer, A.~Parashar, V.~Sarkar, and T.~Krishna,
  ``Understanding reuse, performance, and hardware cost of dnn dataflow: A
  data-centric approach,'' in \emph{Proceedings of the 52nd Annual IEEE/ACM
  International Symposium on Microarchitecture}, 2019, pp. 754--768.

\bibitem{maeri}
H.~Kwon, A.~Samajdar, and T.~Krishna, ``Maeri: Enabling flexible dataflow
  mapping over dnn accelerators via reconfigurable interconnects,'' \emph{ACM
  SIGPLAN Notices}, vol.~53, no.~2, pp. 461--475, 2018.

\bibitem{lane2016deepx}
N.~D. Lane, S.~Bhattacharya, P.~Georgiev, C.~Forlivesi, L.~Jiao, L.~Qendro, and
  F.~Kawsar, ``Deepx: A software accelerator for low-power deep learning
  inference on mobile devices,'' in \emph{2016 15th ACM/IEEE International
  Conference on Information Processing in Sensor Networks (IPSN)}.\hskip 1em
  plus 0.5em minus 0.4em\relax IEEE, 2016, pp. 1--12.

\bibitem{larochelle2007empirical}
H.~Larochelle, D.~Erhan, A.~Courville, J.~Bergstra, and Y.~Bengio, ``An
  empirical evaluation of deep architectures on problems with many factors of
  variation,'' in \emph{Proceedings of the 24th international conference on
  Machine learning}, 2007, pp. 473--480.

\bibitem{li2016high}
H.~Li, X.~Fan, L.~Jiao, W.~Cao, X.~Zhou, and L.~Wang, ``A high performance
  fpga-based accelerator for large-scale convolutional neural networks,'' in
  \emph{2016 26th International Conference on Field Programmable Logic and
  Applications (FPL)}.\hskip 1em plus 0.5em minus 0.4em\relax IEEE, 2016, pp.
  1--9.

\bibitem{ddpg}
T.~P. Lillicrap, J.~J. Hunt, A.~Pritzel, N.~Heess, T.~Erez, Y.~Tassa,
  D.~Silver, and D.~Wierstra, ``Continuous control with deep reinforcement
  learning,'' \emph{arXiv preprint arXiv:1509.02971}, 2015.

\bibitem{lu2019neural}
Q.~Lu, W.~Jiang, X.~Xu, Y.~Shi, and J.~Hu, ``On neural architecture search for
  resource-constrained hardware platforms,'' \emph{arXiv preprint
  arXiv:1911.00105}, 2019.

\bibitem{ma2017optimizing}
Y.~Ma, Y.~Cao, S.~Vrudhula, and J.-s. Seo, ``Optimizing loop operation and
  dataflow in fpga acceleration of deep convolutional neural networks,'' in
  \emph{Proceedings of the 2017 ACM/SIGDA International Symposium on
  Field-Programmable Gate Arrays}, 2017, pp. 45--54.

\bibitem{mirhoseini2017device}
A.~Mirhoseini, H.~Pham, Q.~V. Le, B.~Steiner, R.~Larsen, Y.~Zhou, N.~Kumar,
  M.~Norouzi, S.~Bengio, and J.~Dean, ``Device placement optimization with
  reinforcement learning,'' in \emph{Proceedings of the 34th International
  Conference on Machine Learning-Volume 70}.\hskip 1em plus 0.5em minus
  0.4em\relax JMLR. org, 2017, pp. 2430--2439.

\bibitem{a2c}
V.~Mnih, A.~P. Badia, M.~Mirza, A.~Graves, T.~Lillicrap, T.~Harley, D.~Silver,
  and K.~Kavukcuoglu, ``Asynchronous methods for deep reinforcement learning,''
  in \emph{International conference on machine learning}, 2016, pp. 1928--1937.

\bibitem{mnih2013playing}
V.~Mnih, K.~Kavukcuoglu, D.~Silver, A.~Graves, I.~Antonoglou, D.~Wierstra, and
  M.~Riedmiller, ``Playing atari with deep reinforcement learning,''
  \emph{arXiv preprint arXiv:1312.5602}, 2013.

\bibitem{motamedi2016design}
M.~Motamedi, P.~Gysel, V.~Akella, and S.~Ghiasi, ``Design space exploration of
  fpga-based deep convolutional neural networks,'' in \emph{2016 21st Asia and
  South Pacific Design Automation Conference (ASP-DAC)}.\hskip 1em plus 0.5em
  minus 0.4em\relax IEEE, 2016, pp. 575--580.

\bibitem{novillo2014samplepgo}
D.~Novillo, ``Samplepgo-the power of profile guided optimizations without the
  usability burden,'' in \emph{2014 LLVM Compiler Infrastructure in HPC}.\hskip
  1em plus 0.5em minus 0.4em\relax IEEE, 2014, pp. 22--28.

\bibitem{paliwal2020reinforced}
A.~Paliwal, F.~Gimeno, V.~G. Nair, Y.~Li, M.~Lubin, P.~Kohli, and O.~Vinyals,
  ``Reinforced genetic algorithm learning for optimizing computation graphs,''
  2020.

\bibitem{timeloop}
A.~Parashar, P.~Raina, Y.~S. Shao, Y.-H. Chen, V.~A. Ying, A.~Mukkara,
  R.~Venkatesan, B.~Khailany, S.~W. Keckler, and J.~Emer, ``Timeloop: A
  systematic approach to dnn accelerator evaluation,'' in \emph{2019 IEEE
  International Symposium on Performance Analysis of Systems and Software
  (ISPASS)}.\hskip 1em plus 0.5em minus 0.4em\relax IEEE, 2019, pp. 304--315.

\bibitem{parsa2019pabo}
M.~Parsa, A.~Ankit, A.~Ziabari, and K.~Roy, ``Pabo: Pseudo agent-based
  multi-objective bayesian hyperparameter optimization for efficient neural
  accelerator design,'' \emph{arXiv preprint arXiv:1906.08167}, 2019.

\bibitem{bayes}
M.~Pelikan, D.~E. Goldberg, E.~Cant{\'u}-Paz \emph{et~al.}, ``Boa: The bayesian
  optimization algorithm,'' in \emph{Proceedings of the genetic and
  evolutionary computation conference GECCO-99}, vol.~1, 1999, pp. 525--532.

\bibitem{pinto2009high}
N.~Pinto, D.~Doukhan, J.~J. DiCarlo, and D.~D. Cox, ``A high-throughput
  screening approach to discovering good forms of biologically inspired visual
  representation,'' \emph{PLoS computational biology}, vol.~5, no.~11, 2009.

\bibitem{popov2017data}
I.~Popov, N.~Heess, T.~Lillicrap, R.~Hafner, G.~Barth-Maron, M.~Vecerik,
  T.~Lampe, Y.~Tassa, T.~Erez, and M.~Riedmiller, ``Data-efficient deep
  reinforcement learning for dexterous manipulation,'' \emph{arXiv preprint
  arXiv:1704.03073}, 2017.

\bibitem{gpt2}
A.~Radford, J.~Wu, R.~Child, D.~Luan, D.~Amodei, and I.~Sutskever, ``Language
  models are unsupervised multitask learners,'' \emph{OpenAI Blog}, vol.~1,
  no.~8, p.~9, 2019.

\bibitem{ragan2013halide}
J.~Ragan-Kelley, C.~Barnes, A.~Adams, S.~Paris, F.~Durand, and S.~Amarasinghe,
  ``Halide: a language and compiler for optimizing parallelism, locality, and
  recomputation in image processing pipelines,'' \emph{Acm Sigplan Notices},
  vol.~48, no.~6, pp. 519--530, 2013.

\bibitem{reagen2017case}
B.~Reagen, J.~M. Hern{\'a}ndez-Lobato, R.~Adolf, M.~Gelbart, P.~Whatmough,
  G.-Y. Wei, and D.~Brooks, ``A case for efficient accelerator design space
  exploration via bayesian optimization,'' in \emph{2017 IEEE/ACM International
  Symposium on Low Power Electronics and Design (ISLPED)}.\hskip 1em plus 0.5em
  minus 0.4em\relax IEEE, 2017, pp. 1--6.

\bibitem{minerva}
B.~Reagen, P.~Whatmough, R.~Adolf, S.~Rama, H.~Lee, S.~K. Lee, J.~M.
  Hern{\'a}ndez-Lobato, G.-Y. Wei, and D.~Brooks, ``Minerva: Enabling
  low-power, highly-accurate deep neural network accelerators,'' in \emph{2016
  ACM/IEEE 43rd Annual International Symposium on Computer Architecture
  (ISCA)}.\hskip 1em plus 0.5em minus 0.4em\relax IEEE, 2016, pp. 267--278.

\bibitem{salimans2017evolution}
T.~Salimans, J.~Ho, X.~Chen, S.~Sidor, and I.~Sutskever, ``Evolution strategies
  as a scalable alternative to reinforcement learning,'' \emph{arXiv preprint
  arXiv:1703.03864}, 2017.

\bibitem{sandler2018mobilenetv2}
M.~Sandler, A.~Howard, M.~Zhu, A.~Zhmoginov, and L.-C. Chen, ``Mobilenetv2:
  Inverted residuals and linear bottlenecks,'' in \emph{Proceedings of the IEEE
  conference on computer vision and pattern recognition}, 2018, pp. 4510--4520.

\bibitem{sankaradas2009massively}
M.~Sankaradas, V.~Jakkula, S.~Cadambi, S.~Chakradhar, I.~Durdanovic,
  E.~Cosatto, and H.~P. Graf, ``A massively parallel coprocessor for
  convolutional neural networks,'' in \emph{2009 20th IEEE International
  Conference on Application-specific Systems, Architectures and
  Processors}.\hskip 1em plus 0.5em minus 0.4em\relax IEEE, 2009, pp. 53--60.

\bibitem{santoro2018energy}
G.~Santoro, M.~R. Casu, V.~Peluso, A.~Calimera, and M.~Alioto,
  ``Energy-performance design exploration of a low-power microprogrammed
  deep-learning accelerator,'' in \emph{2018 Design, Automation \& Test in
  Europe Conference \& Exhibition (DATE)}.\hskip 1em plus 0.5em minus
  0.4em\relax IEEE, 2018, pp. 1151--1154.

\bibitem{schkufza2013stochastic}
E.~Schkufza, R.~Sharma, and A.~Aiken, ``Stochastic superoptimization,''
  \emph{ACM SIGARCH Computer Architecture News}, vol.~41, no.~1, pp. 305--316,
  2013.

\bibitem{ppo2}
J.~Schulman, F.~Wolski, P.~Dhariwal, A.~Radford, and O.~Klimov, ``Proximal
  policy optimization algorithms,'' \emph{arXiv preprint arXiv:1707.06347},
  2017.

\bibitem{aladdin}
Y.~S. Shao, S.~L. Xi, V.~Srinivasan, G.-Y. Wei, and D.~Brooks, ``Co-designing
  accelerators and soc interfaces using gem5-aladdin,'' in \emph{2016 49th
  Annual IEEE/ACM International Symposium on Microarchitecture (MICRO)}.\hskip
  1em plus 0.5em minus 0.4em\relax IEEE, 2016, pp. 1--12.

\bibitem{shen2017maximizing}
Y.~Shen, M.~Ferdman, and P.~Milder, ``Maximizing cnn accelerator efficiency
  through resource partitioning,'' in \emph{2017 ACM/IEEE 44th Annual
  International Symposium on Computer Architecture (ISCA)}.\hskip 1em plus
  0.5em minus 0.4em\relax IEEE, 2017, pp. 535--547.

\bibitem{stamoulis2018hyperpower}
D.~Stamoulis, E.~Cai, D.-C. Juan, and D.~Marculescu, ``Hyperpower: Power-and
  memory-constrained hyper-parameter optimization for neural networks,'' in
  \emph{2018 Design, Automation \& Test in Europe Conference \& Exhibition
  (DATE)}.\hskip 1em plus 0.5em minus 0.4em\relax IEEE, 2018, pp. 19--24.

\bibitem{stamoulis2018designing}
D.~Stamoulis, T.-W. Chin, A.~K. Prakash, H.~Fang, S.~Sajja, M.~Bognar, and
  D.~Marculescu, ``Designing adaptive neural networks for energy-constrained
  image classification,'' in \emph{Proceedings of the International Conference
  on Computer-Aided Design}, 2018, pp. 1--8.

\bibitem{steuwer2017lift}
M.~Steuwer, T.~Remmelg, and C.~Dubach, ``Lift: a functional data-parallel ir
  for high-performance gpu code generation,'' in \emph{2017 IEEE/ACM
  International Symposium on Code Generation and Optimization (CGO)}.\hskip 1em
  plus 0.5em minus 0.4em\relax IEEE, 2017, pp. 74--85.

\bibitem{sutskever2014sequence}
I.~Sutskever, O.~Vinyals, and Q.~V. Le, ``Sequence to sequence learning with
  neural networks,'' in \emph{Advances in neural information processing
  systems}, 2014, pp. 3104--3112.

\bibitem{reinforce}
R.~S. Sutton, D.~A. McAllester, S.~P. Singh, and Y.~Mansour, ``Policy gradient
  methods for reinforcement learning with function approximation,'' in
  \emph{Advances in neural information processing systems}, 2000, pp.
  1057--1063.

\bibitem{szegedy2017inception}
C.~Szegedy, S.~Ioffe, V.~Vanhoucke, and A.~A. Alemi, ``Inception-v4,
  inception-resnet and the impact of residual connections on learning,'' in
  \emph{Thirty-first AAAI conference on artificial intelligence}, 2017.

\bibitem{tan2019mnasnet}
M.~Tan, B.~Chen, R.~Pang, V.~Vasudevan, M.~Sandler, A.~Howard, and Q.~V. Le,
  ``Mnasnet: Platform-aware neural architecture search for mobile,'' in
  \emph{Proceedings of the IEEE Conference on Computer Vision and Pattern
  Recognition}, 2019, pp. 2820--2828.

\bibitem{mnasnet}
M.~Tan, B.~Chen, R.~Pang, V.~Vasudevan, M.~Sandler, A.~Howard, and Q.~V. Le,
  ``Mnasnet: Platform-aware neural architecture search for mobile,'' in
  \emph{Proceedings of the IEEE Conference on Computer Vision and Pattern
  Recognition}, 2019, pp. 2820--2828.

\bibitem{tan2019efficientnet}
M.~Tan and Q.~V. Le, ``Efficientnet: Rethinking model scaling for convolutional
  neural networks,'' \emph{arXiv preprint arXiv:1905.11946}, 2019.

\bibitem{torrado2018deep}
R.~R. Torrado, P.~Bontrager, J.~Togelius, J.~Liu, and D.~Perez-Liebana, ``Deep
  reinforcement learning for general video game ai,'' in \emph{2018 IEEE
  Conference on Computational Intelligence and Games (CIG)}.\hskip 1em plus
  0.5em minus 0.4em\relax IEEE, 2018, pp. 1--8.

\bibitem{vasilache2018tensor}
N.~Vasilache, O.~Zinenko, T.~Theodoridis, P.~Goyal, Z.~DeVito, W.~S. Moses,
  S.~Verdoolaege, A.~Adams, and A.~Cohen, ``Tensor comprehensions:
  Framework-agnostic high-performance machine learning abstractions,''
  \emph{arXiv preprint arXiv:1802.04730}, 2018.

\bibitem{vaswani2017attention}
A.~Vaswani, N.~Shazeer, N.~Parmar, J.~Uszkoreit, L.~Jones, A.~N. Gomez,
  {\L}.~Kaiser, and I.~Polosukhin, ``Attention is all you need,'' in
  \emph{Advances in neural information processing systems}, 2017, pp.
  5998--6008.

\bibitem{wei2017dlvm}
R.~Wei, L.~Schwartz, and V.~Adve, ``Dlvm: A modern compiler infrastructure for
  deep learning systems,'' \emph{arXiv preprint arXiv:1711.03016}, 2017.

\bibitem{wei2018tgpa}
X.~Wei, Y.~Liang, X.~Li, C.~H. Yu, P.~Zhang, and J.~Cong, ``Tgpa: tile-grained
  pipeline architecture for low latency cnn inference,'' in \emph{Proceedings
  of the International Conference on Computer-Aided Design}, 2018, pp. 1--8.

\bibitem{wei2017automated}
X.~Wei, C.~H. Yu, P.~Zhang, Y.~Chen, Y.~Wang, H.~Hu, Y.~Liang, and J.~Cong,
  ``Automated systolic array architecture synthesis for high throughput cnn
  inference on fpgas,'' in \emph{Proceedings of the 54th Annual Design
  Automation Conference 2017}, 2017, pp. 1--6.

\bibitem{whaley1998automatically}
R.~C. Whaley and J.~J. Dongarra, ``Automatically tuned linear algebra
  software,'' in \emph{SC'98: Proceedings of the 1998 ACM/IEEE conference on
  Supercomputing}.\hskip 1em plus 0.5em minus 0.4em\relax IEEE, 1998, pp.
  38--38.

\bibitem{gnmt}
Y.~Wu, M.~Schuster, Z.~Chen, Q.~V. Le, M.~Norouzi, W.~Macherey, M.~Krikun,
  Y.~Cao, Q.~Gao, K.~Macherey \emph{et~al.}, ``Google's neural machine
  translation system: Bridging the gap between human and machine translation,''
  \emph{arXiv preprint arXiv:1609.08144}, 2016.

\bibitem{acktr}
Y.~Wu, E.~Mansimov, R.~B. Grosse, S.~Liao, and J.~Ba, ``Scalable trust-region
  method for deep reinforcement learning using kronecker-factored
  approximation,'' in \emph{Advances in neural information processing systems},
  2017, pp. 5279--5288.

\bibitem{wu2016training}
Y.~Wu and Y.~Tian, ``Training agent for first-person shooter game with
  actor-critic curriculum learning,'' 2016.

\bibitem{yang2020co}
L.~Yang, Z.~Yan, M.~Li, H.~Kwon, L.~Lai, T.~Krishna, V.~Chandra, W.~Jiang, and
  Y.~Shi, ``Co-exploration of neural architectures and heterogeneous asic
  accelerator designs targeting multiple tasks,'' \emph{arXiv preprint
  arXiv:2002.04116}, 2020.

\bibitem{yang2018dnn}
X.~Yang, M.~Gao, J.~Pu, A.~Nayak, Q.~Liu, S.~E. Bell, J.~O. Setter, K.~Cao,
  H.~Ha, C.~Kozyrakis \emph{et~al.}, ``Dnn dataflow choice is overrated,''
  \emph{arXiv preprint arXiv:1809.04070}, 2018.

\bibitem{yin2018141}
S.~Yin, P.~Ouyang, S.~Zheng, D.~Song, X.~Li, L.~Liu, and S.~Wei, ``A 141 uw,
  2.46 pj/neuron binarized convolutional neural network based self-learning
  speech recognition processor in 28nm cmos,'' in \emph{2018 IEEE Symposium on
  VLSI Circuits}.\hskip 1em plus 0.5em minus 0.4em\relax IEEE, 2018, pp.
  139--140.

\bibitem{yonghui2016bridging}
W.~Yonghui, M.~Schuster, Z.~Chen, Q.~Le, M.~Norouzi, W.~Macherey, M.~Krikun,
  Y.~Cao, Q.~Gao, K.~Macherey \emph{et~al.}, ``Bridging the gap between human
  and machine translation,'' \emph{arXiv preprint arXiv:1609.08144}, 2016.

\bibitem{cong_fpga}
C.~Zhang, P.~Li, G.~Sun, Y.~Guan, B.~Xiao, and J.~Cong, ``Optimizing fpga-based
  accelerator design for deep convolutional neural networks,'' in
  \emph{Proceedings of the 2015 ACM/SIGDA International Symposium on
  Field-Programmable Gate Arrays}, 2015, pp. 161--170.

\bibitem{zhang2016energy}
C.~Zhang, D.~Wu, J.~Sun, G.~Sun, G.~Luo, and J.~Cong, ``Energy-efficient cnn
  implementation on a deeply pipelined fpga cluster,'' in \emph{Proceedings of
  the 2016 International Symposium on Low Power Electronics and Design}, 2016,
  pp. 326--331.

\bibitem{zhang2018dnnbuilder}
X.~Zhang, J.~Wang, C.~Zhu, Y.~Lin, J.~Xiong, W.-m. Hwu, and D.~Chen,
  ``Dnnbuilder: an automated tool for building high-performance dnn hardware
  accelerators for fpgas,'' in \emph{2018 IEEE/ACM International Conference on
  Computer-Aided Design (ICCAD)}.\hskip 1em plus 0.5em minus 0.4em\relax IEEE,
  2018, pp. 1--8.

\bibitem{mrna}
Z.~Zhao, H.~Kwon, S.~Kuhar, W.~Sheng, Z.~Mao, and T.~Krishna, ``mrna: Enabling
  efficient mapping space exploration for a reconfiguration neural
  accelerator,'' in \emph{2019 IEEE International Symposium on Performance
  Analysis of Systems and Software (ISPASS)}.\hskip 1em plus 0.5em minus
  0.4em\relax IEEE, 2019, pp. 282--292.

\bibitem{zheng2019ultra}
S.~Zheng, P.~Ouyang, D.~Song, X.~Li, L.~Liu, S.~Wei, and S.~Yin, ``An ultra-low
  power binarized convolutional neural network-based speech recognition
  processor with on-chip self-learning,'' \emph{IEEE Transactions on Circuits
  and Systems I: Regular Papers}, vol.~66, no.~12, pp. 4648--4661, 2019.

\bibitem{zhong2017design}
G.~Zhong, A.~Prakash, S.~Wang, Y.~Liang, T.~Mitra, and S.~Niar, ``Design space
  exploration of fpga-based accelerators with multi-level parallelism,'' in
  \emph{Design, Automation \& Test in Europe Conference \& Exhibition (DATE),
  2017}.\hskip 1em plus 0.5em minus 0.4em\relax IEEE, 2017, pp. 1141--1146.

\bibitem{zhong2009tuning}
S.~Zhong, Y.~Shen, and F.~Hao, ``Tuning compiler optimization options via
  simulated annealing,'' in \emph{2009 Second International Conference on
  Future Information Technology and Management Engineering}.\hskip 1em plus
  0.5em minus 0.4em\relax IEEE, 2009, pp. 305--308.

\end{thebibliography}
